\documentclass[journal]{article}
\usepackage[utf8]{inputenc}
\usepackage{graphicx}
\usepackage[caption = false]{subfig}
\usepackage[utf8]{inputenc}		
\usepackage{empheq}						
\usepackage{paralist}					
\usepackage{amsmath,amsfonts,amssymb,mathtools}
\usepackage{bbold}						
\usepackage{float}
\usepackage{hyperref}					
\usepackage{todonotes}
\usepackage{gensymb}
\usetikzlibrary{arrows,decorations,backgrounds,patterns,matrix,shapes,fit,calc,shadows,plotmarks,pgfplots.groupplots}
\title{Longitudinal charge distribution measurement of non-relativistic ion beams using coherent transition radiation}
\author{R. Singh\thanks{r.singh@gsi.de}
\textsuperscript{1}, T. Reichert\textsuperscript{1} \\
	    \textsuperscript{1}GSI Helmholtzzentrum f\"ur Schwerionenforschung GmbH,\\ Darmstadt, Germany  \\}
\date{June 2021}

\begin{document}

\maketitle
\section{Abstract}
Longitudinal charge profile measurements or bunch shape measurements is a challenge for temporally short non-relativistic bunches. The field profile has a larger longitudinal extent compared to the charge profile for such beams. This affects the ability of field sensing devices such as phase pick-ups or wall current monitors to measure charge distribution. Here we evaluate the feasibility for usage of coherent transition~\cite{Ginzburg1,TM} and diffraction radiation from non-relativistic beams for bunch shape measurements.
\section{Introduction}
 Charge profile measurements are essential for verification of LINAC beam dynamics, commissioning and optimization of LINACs. This has been historically challenging for non-relativistic beams since the non-destructive field monitors could not be used for this purpose. Meanwhile there are other devices such as fast Faraday cup designs~\cite{BARC_Jose,Fermilab_Shemyakin} (FFCs) which use the ground plates to the shield the detection plate to avoid early arrival of beam field. However, 
the signal generation process including the role of secondary electron emission is not fully understood . Further, Faraday cups are interceptive measurements which makes their usage unattractive with high intensity beams as well as unfeasible for phase space measurements. The commonly used alternative is the bunch shape monitor~\cite{Feschenko} or "Feschenko monitor", which relies on electron emission from a wire after the beam interaction. This device is however limited to a temporally averaged measurement due to small interaction of electron emitting wire with beam and inherently relies on stable beam conditions. Therefore such averaged measurements can be misleading when there are shot- to-shot current fluctuations or any other effects which lead to non-reproducible charge distribution in the duration of measurements. Interestingly, there is no bench-marking of the aforementioned devices available in literature. 

In this report, we study a non-destructive bunch-by-bunch longitudinal charge profile measurement alternative based on coherent transition~\cite{Ginzburg1,TM} and diffraction radiation. Transition radiation from its very nature provides a measure of relative charge variation. The range of interest in terms of beam velocities is $\beta \in [0.05,0.9]$ and bunch lengths between 50-500 ps (1$\sigma$). In the first section, we will discuss the generation process, angular distribution of the radiated field, formation zone, target size dependence as well as provide signal estimates using available analytical models for transition radiation (TR). The non-destructiveness of the measurement principle by means of a hole in the target will also be discussed. In the second section, we compare the analytical results with 
electromagnetic particle-in-cell simulations.  Few examples for relativistic beams i.e. $\beta = 0.99$ are shown to connect the results in this report to the electron beam examples found extensively in literature as well as to discuss the feasibility of our proposal for relativistic beams. In the final section, first results from a prototype installed at GSI UNILAC are shown.

\section{Analytical estimates}
\subsection{Electric field calculations with ideal set-up}
\begin{figure}[H]
\centering  
    \includegraphics[width=0.8\textwidth]{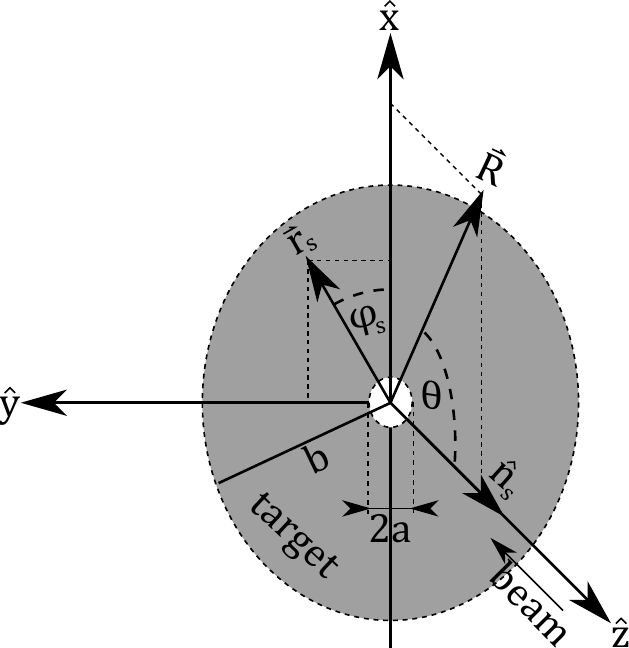}
    \caption{Coordinate system and the target plane}
    \label{fig:1}
\end{figure}

Figure~\ref{fig:1} shows a potential set-up for the proposed bunch shape measurement using transition radiation. A perfectly conducting ring-shaped target (inner/outer radius $a$ , $b$ respectively) is located at $z=0$ so that its surface normal is given by $\vec{n}_s = (0,0,1)^T$. The charged particle beam originates from infinity and travels through vacuum approaching the target with perpendicular incidence  with velocity $\vec{v}_e = -\beta c_0 \hat{e}_z$. The target can be made of 
standard metals since they can practically all be considered perfect electric conductors (PEC) in the frequency regime of interest ($\leq 30$ GHz). The hole of radius $a$ in the target is for non-destructive measurements.  

 As depicted in Fig.\ref{fig:1}, the field monitor or observation point is located in the $x-z$ plane sensitive to radiation emitted in that plane $k_x\hat{e}_x+k_z\hat{e}_z$.
\begin{align}
    \vec{R} =\left( \begin{matrix} x \\ 0 \\ z \end{matrix} \right) = \left( \begin{matrix}  R \sin(\theta) \\ 0 \\ R \cos(\theta) \end{matrix} \right) \label{eq1}
\end{align}
and the co-ordinates on the target plate are given by,
\begin{align}
    \vec{r}_s =\left( \begin{matrix} x_s \\ y_s \\ 0 \end{matrix} \right) = \left( \begin{matrix}  r_s \cos(\phi_s) \\ r_s \sin(\phi_s) \\0 \end{matrix} \right) \label{eq2}
\end{align}
A comprehensive and modern review of the generation process of transition radiation is given in~\cite{Fiorito} which is a pre-requisite for the reader of this report. In this report, we make an attempt to use the same formalism as~\cite{Fiorito} and restrict ourselves to the relevant and special case of normal incidence of beam on the target. The Fourier transform of the "source" field of the moving charged particle
is given as, 
\begin{align}
    E_{s,x}(z=0,\omega) &= G(r_s,\omega) \cos(\phi_s) \nonumber \\[0.15cm] 
    E_{s,y}(z=0,\omega) &= G(r_s,\omega) \sin(\phi_s) 
    \nonumber \\[0.15cm]
    \text{with} \ \   G(r_s) &:= \frac{e \alpha}{ 4\pi\epsilon_0 \pi \beta c} K_1 (\alpha r_s) 
    \label{eq3}
\end{align}
where $K_1$ represents the modified Bessel's function of the second kind and $\alpha = \omega/{\beta\gamma c}$.  
In~\cite{Fiorito} the scattered/radiated transition radiation field originates from a virtual magnetic surface current induced on the target due to the source field,
\begin{align}
    \vec{j}_{vm} &= \frac{c}{4\pi} \hat{n}_s \times \vec{E}_s \nonumber \\
    \Rightarrow \vec{j}_{vm} &= \frac{c}{4\pi} \left( \begin{matrix} -E_{s,y} \\ E_{s,x} \\ 0 \end{matrix} \right)
    \label{eq4}
\end{align}
As evident in Eq.~\ref{eq4}, for the case of normal incidence, only the transverse components $E_{s,x}$ and $E_{s,y}$ are of relevance in formation of virtual magnetic surface current.
The associated vector potential can be calculated by inserting \eqref{eq3} and \eqref{eq4} in Eq.(19) of~\cite{Fiorito},
\begin{align}
    \vec{A} &= \frac{2}{c} \int_a^b dr_s \int_0^{2\pi} d\phi_s r_s \vec{j}_{vm} \frac{\exp(i k R_s)}{R_s}  \nonumber \\[0.15cm]
    &= \frac{1}{2\pi} \int_a^b dr_s \int_0^{2\pi} d\phi_s r_s G(r_s) \frac{\exp(i k R_s)}{R_s} \left( \begin{matrix} -\sin(\phi_s) \\ \cos(\phi_s) \\ 0 \end{matrix} \right)  \label{eq5}
\end{align}
where $R_s$ is the norm of the difference vector between $\vec{R}$ and $\vec{r}_s$
\begin{align}
    R_s &= \sqrt{(x-x_s)^2 + y_s^2 + z^2} \nonumber \\[0.15cm]
    &= \sqrt{x^2 + z^2 + r_s^2 - 2 x r_s \cos(\phi_s)} \label{eq6}
\end{align}

Since $R_s$ is an even function of $\phi_s$ 
 and gets multiplied by an odd function in $A_x$ the symmetric integration over
$\phi_s$
vanishes. Analogously the same integration from $A_y$ can be limited to the interval $[0,\pi]$ with a doubled integrand resulting in,

\begin{align}
    \vec{A} &= -\frac{\hat{e}_y}{\pi} \int_a^b dr_s \int_0^{\pi}d\phi_s r_s \cos(\phi_s) G(r_s) \frac{\exp(i k R_s)}{R_s}   \label{eq10}
\end{align}
It should be noted that Eq.~\ref{eq10} implies that only the source fields in the plane of observation contribute to the vector potential perpendicular to it. From the vector potential, the electric fields can be obtained by,
\begin{align}
    \vec{E} = - \nabla \times \vec{A}
    \label{eq11}
\end{align}
Thus the generated transition radiation field in the plane of observation is the result of source fields only in that same plane. A quasi-spherical approximation (QSA) for calculation of electric fields is made ($R_s \approx R$ and $1/kR \ll 1$)in~\cite{Fiorito} due to reasons to computational complexity, i.e.
$    \vec{E} = -i\vec{k} \times \vec{A} $
 and the validity of this QSA is discussed and quantified. For our purpose of bunch length measurements, the target size, wavelength of interest and monitor distance are of the same order and the validity of quasi-spherical approximation is not clear. Therefore, we calculate the radiated electric field without the quasi-spherical approximation. As there is only a $y$-component in $\vec{A}$ we can write,
\begin{align}
    \vec{E} = - \nabla \times \vec{A} = \left(\begin{matrix} \partial_z A_y \\ 0 \\ -\partial_x A_y \end{matrix} \right) \nonumber 
\end{align}

For this purpose we show
\begin{align}
    \partial_z \frac{\exp(i k \tilde{R_s})}{\tilde{R_s}} 
    &= \frac{z \exp(i k \tilde{R_s})}{\tilde{R_s}^3}\left(i k \tilde{R_s} - 1\right) \nonumber 
\end{align}
and 
\begin{align}
    \partial_x \frac{\exp(i k \tilde{R_s})}{\tilde{R_s}} 
    &= \frac{(x+r_s\cos(\tilde{\phi_s})) \exp(i k \tilde{R_s})}{\tilde{R_s}^3}\left(i k \tilde{R_s} - 1\right) \label{eq13}
\end{align}
and thus we arrive at,
\begin{align}
\vec{E} 
&= \frac{1}{\pi} \int_a^b dr_s \int_0^\pi d\tilde{\phi_s} G(\alpha,r_s) r_s \cos(\tilde{\phi_s}) \frac{\exp[i k \tilde{R_s}]}{\tilde{R_s}^3} (1-i k \tilde{R_s}) \left( \begin{matrix} z \\ 0 \\ -x - r_s \cos(\tilde{\phi_s})  \end{matrix}\right) \nonumber \\[0.2cm] 
\label{eq14}
\end{align}
Inserting~Eq.\eqref{eq3} we have the expression for the special case of normal incidence radiation (NIR) which should be valid at any position from the target,
\begin{align}
    \vec{E} = \frac{2 e}{(4\pi\epsilon_0)\pi  \beta^2 \gamma c \lambda } \int_a^b dr_s \int_0^\pi d\tilde{\phi_s} K_1(\frac{2 \pi r_s}{\beta\gamma\lambda}) r_s \cos(\tilde{\phi_s}) \frac{\exp[i k \tilde{R_s}]}{\tilde{R_s}^3} (1-i k \tilde{R_s}) \left( \begin{matrix} z \\ 0 \\ -x - r_s \cos(\tilde{\phi_s})  \end{matrix}\right) \nonumber \\[0.2cm] 
\label{eq15}
\end{align}
It is possible to further simplify the result in Eq.~\eqref{eq15} to a single integral in case of quasi-spherical approximation as discussed in the Appendix. 

\begin{figure}[H]
\centering 
\subfloat[]{\includegraphics[width=0.48\textwidth]{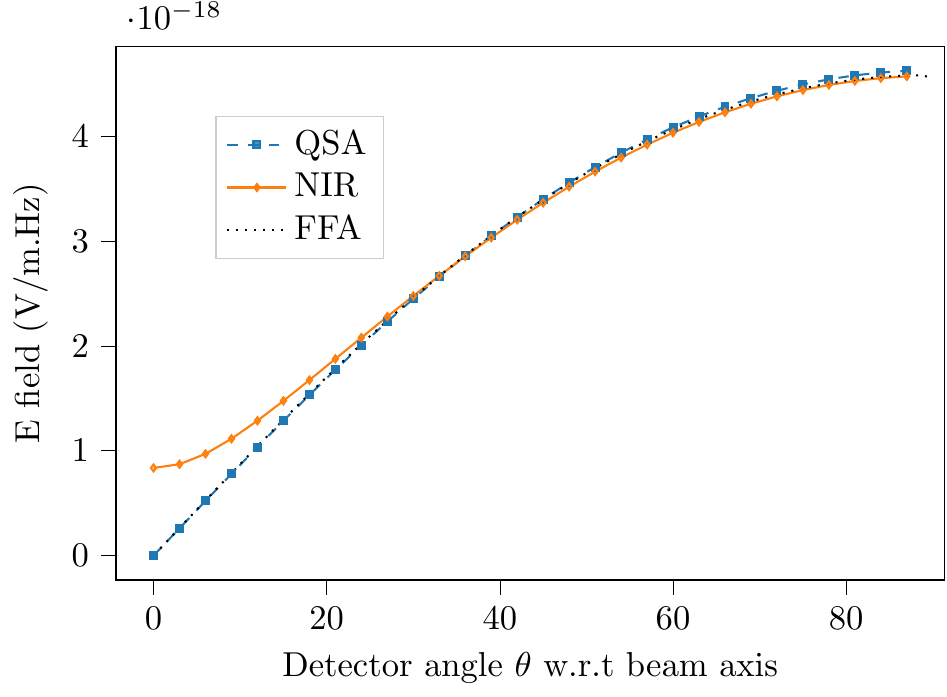}} 
\subfloat[]{\includegraphics[width=0.48\textwidth]{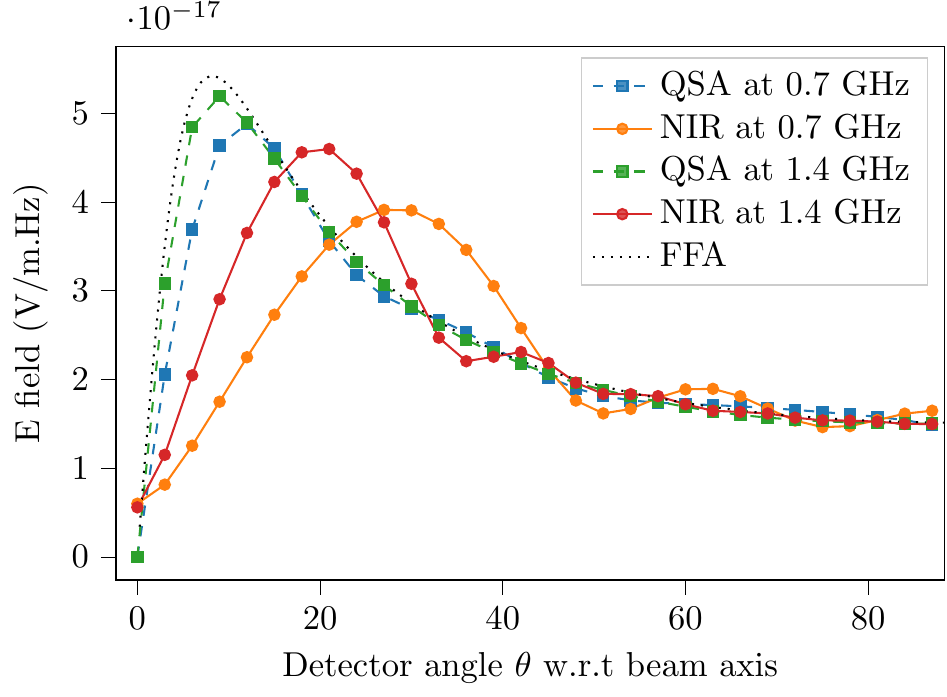}}
\caption{ (a) Comparison of Normal incident radiation (NIR), quasi-spherical approximation (QSA) and far field approximation (FFA) for $\beta = 0.3$ for target radius $b$ = 0.4 m, monitor distance $R$ = 0.4 m and frequency = 0.7 GHz. (b) Comparison of NIR, QSA and FFA for target radius $b$ = 1.2 m, monitor distance $R$= 1.2 m at $\beta = 0.99$.}
    \label{fig:2}
\end{figure}
In Fig.~\ref{fig:2} we compare the transition radiation angular distribution of three cases, the traditional far field result (FFA)~\cite{TM}, quasi spherical approximation (QSA)~\cite{Fiorito} and the "exact" estimate for the special case of normal incidence (NIR) shown in Eq.~\ref{eq15}. There is a significant discrepancy between the QSA and NIR for lower frequencies and distances of interest especially for smaller detection angles $\theta$. Based on this observation, we will utilize NIR as the basis for our analytical transition radiation field calculations in this report.

\subsection{Target size dependence}
\begin{figure}[H]
\centering 
\subfloat[]{\includegraphics[width=0.48\textwidth]{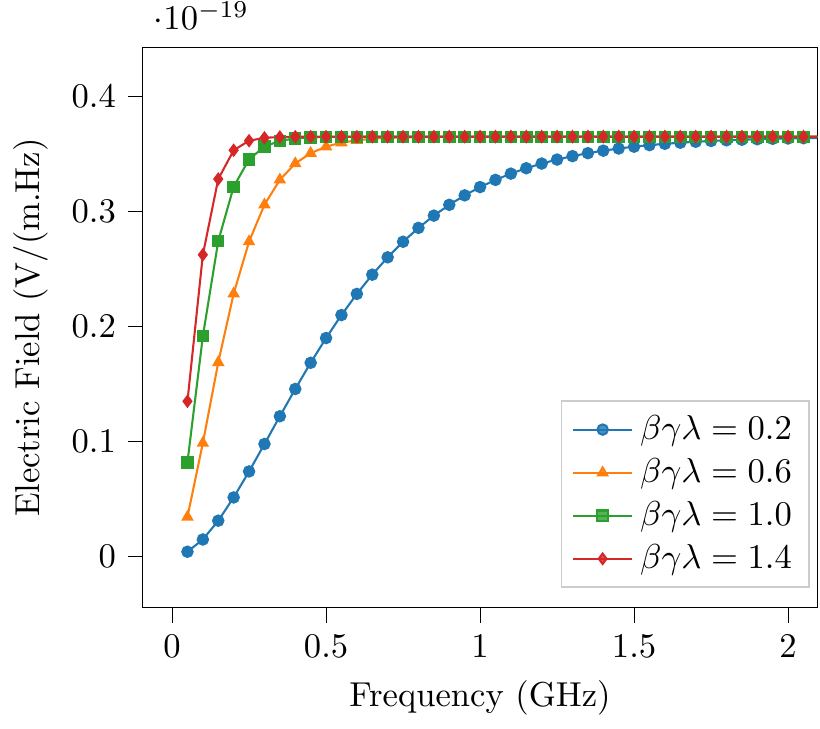}} 
\subfloat[]{\includegraphics[width=0.48\textwidth]{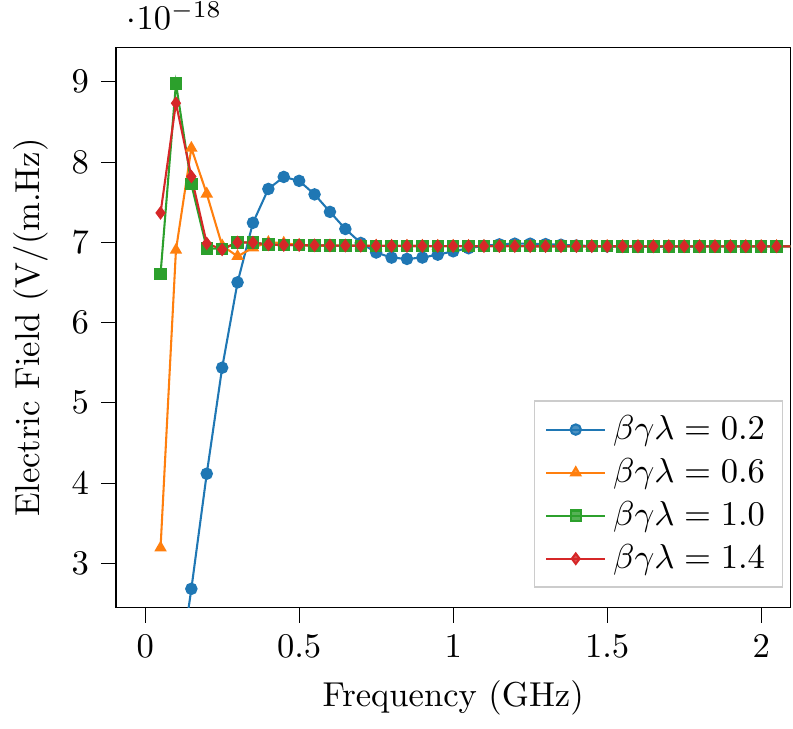}}
\caption{ (a) Frequency response as a function of target size for $\beta = 0.05$ at monitor distance $R$ = 2.0 m (b) Frequency response as a function of target size for $\beta = 0.9$ at monitor distance $R$ = 2.0 m.}
\label{fig:3}
\end{figure}

\begin{figure}[H]
\centering  
    \includegraphics[width=0.8\textwidth]{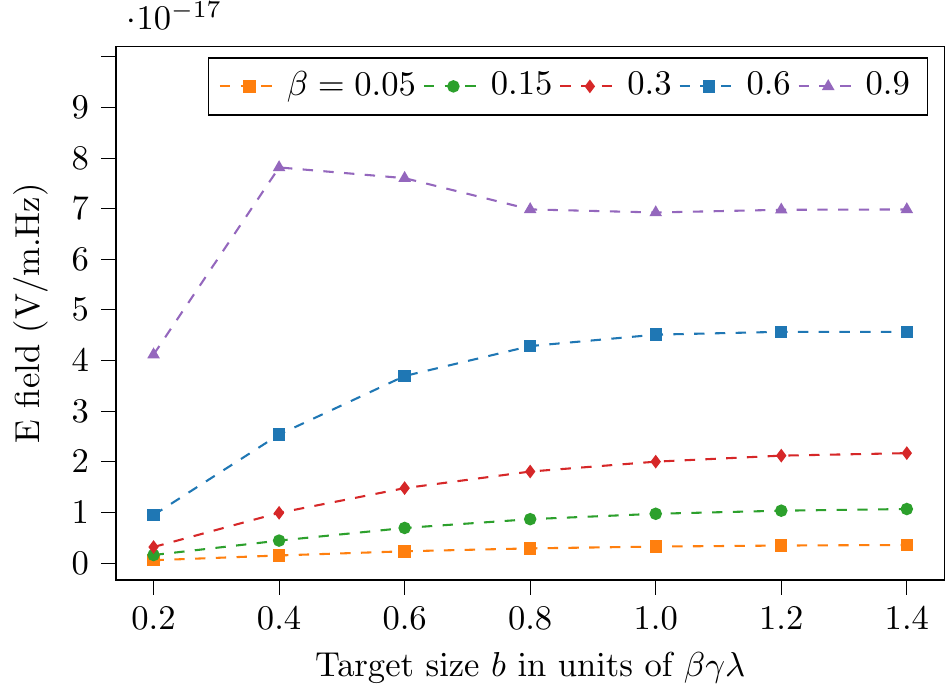}
    \caption{Electric Field spectral amplitude dependence on target size.}
    \label{fig:4}
\end{figure}

It is well known from literature~\cite{Verzilov,Dobrovolsky_Shulga,Fiorito}, that the target size should be larger than the transverse source size in order to avoid deformations in the generated transition radiation fields with respect to the far field angular distribution. The effective source field extent is given by $r_\text{eff} = \beta\gamma\lambda$ for generation of radiation at wavelength $\lambda$. Beyond this radius the incident field's contribution to the radiation field becomes very small. This dependence can be readily seen in the argument of modified Bessel's function shown in Eq.~\ref{eq3}.  A finite target size thus introduces a "high pass filter" behavior at a frequency $f_{cut}$  $\approx\frac{b}{\beta\gamma c}$ where $b$ is the target radius. The calculated frequency responses are shown in Fig.~\ref{fig:3} for various target sizes in units of $\beta\gamma\lambda$ for a monitor placed at the angle of 60\degree at a distance of 2 m for two beam velocities. The $\lambda$ chosen is 1 m corresponding to 300 MHz in vacuum. The high pass behavior in the frequency response is evident in the plots.
Figure~\ref{fig:4} shows the electric field amplitude at a detector angle of $\theta=70\degree$ for different particle velocities as a function of target size. The field reaches the nominal far field values for $\beta\gamma\lambda>0.8$. This could be used as a rule of thumb for determining the smallest target size required to avoid target size dependence on the radiation for a given reference wavelength $\lambda_{ref}$.

\subsection{Formation zone and monitor placement}
Ginzburg and Tsytovich~\cite{Ginzburg1} reviewed theoretical aspects of the transition radiation phenomenon
discussed the concept of radiation formation zone. 
It is defined as the distance travelled by the charged particle while the radiation "emitted" by the charged particle between the start and end points accumulates a phase larger than $\pi$.
This can be understood as the distance travelled by the charged particle where any remnants of the preceding radiation due to interaction with the surroundings is separated. In other words, the charge requires a certain "formation" length after its interaction with a medium to reach its equilibrium or "proper" field again. For the forward transition radiation, this effect is evident as the distance required for the charged particle "proper" coulomb field or "direct field" to separate from the radiated transition radiation. For the backward transition radiation, the question of proper coulomb field and transition radiation does not arise, however there are other practical constraints. Typically it is assumed that charged particle is travelling from infinity towards the target which is not true for most practical cases. Therefore the remnant of the interaction with the boundary (e.g. a beam pipe) which the charge exits will still interfere with the backward transition radiation if a certain minimum separation is unavailable between the boundary and the target. Already from this simple argumentation, it is clear that the formation zone will be a function of observation angle $\theta$.
The formation zone for forward radiation when traversing from a medium to vacuum was derived by Garibian~\cite{Garibian} 
and is given as, 
\begin{align} L_{formation}=\frac{\lambda\beta}{2\pi (1-\beta\cos{\theta})}
\label{eq:formation_length}
\end{align}
. For relativistic beams, the relevant "formation length" is defined with respect to the angle where peak power is radiated, $\theta \approx \frac{1}{\gamma}$ and it can be shown to reduce to this simple widely used expression $L_{formation}=\gamma^2\lambda/2\pi$ for the radiated wavelength $\lambda$. The detector distance $r$ should be much greater than $L_{formation}$, i.e. $r>> L_{formation}$ to observe the far field radiation. Formation length plays a fundamental role in radiator design for high energy physics detectors~\cite{LukePRL}.

The discussion on radiation formation zone was furthered by Verzilov~\cite{Verzilov}, where it was shown that 
the effective transverse size of the incident field $r_\text{eff}$ is linked to a characteristic distance around the target referred to as "prewave" zone. Therein the source has to be considered distributed continuously over the target plane and thus the radiation field will exhibit interference patterns resulting from contributions due to different parts of the target being lit by the incident field.
The qualitative arguments given in ~\cite{Verzilov} are for the small angle approximation and the extent of prewave zone for a beam energy corresponding to a $\gamma$ at the peak of TR again turns out to be $\gamma^2\lambda$. Again, in Fig. 3~\cite{Verzilov} it is shown that the interferences are stronger towards smaller angles and the transition radiation peak is shifted in the prewave zone while these interference effects subside for large observation angles. Both of the aforementioned effects appear different in their origin, yet yield a similar tendency for transition radiation angular distribution as the function of beam parameters. The analytical results discussed in~\cite{Verzilov} only addresses the source size effects and do not consider the "Ginzburg" formation zone mentioned in~\cite{Garibian}. A detailed account of these formation zones and its implications for different detector sizes is given in~\cite{Dobrovolsky_Shulga} and experimentally verified in~\cite{Kalinin}. The Ginzburg formation zone for backward transition radiation is partially addressed in~\cite{Nause} via line diffraction sources however the authors of this report could not reproduce the results due to numerical difficulties. In this section, we only consider the Verzilov prewave zone in context of backward transition radiation and show the angular distributions calculated using NIR (Eq.~\ref{eq15}) and its deviation from the far field distribution. With reference to the discussion of detector size in~\cite{Dobrovolsky_Shulga}, our detectors can be understood as arbitrarily small or "dotted".  We will resort to full electromagnetic simulations in the next section which should include the effects of both Ginzburg "formation zone" and Verzilov "prewave zone". 

\begin{figure}[H]
\centering 
\subfloat[]{\includegraphics[width=0.48\textwidth]{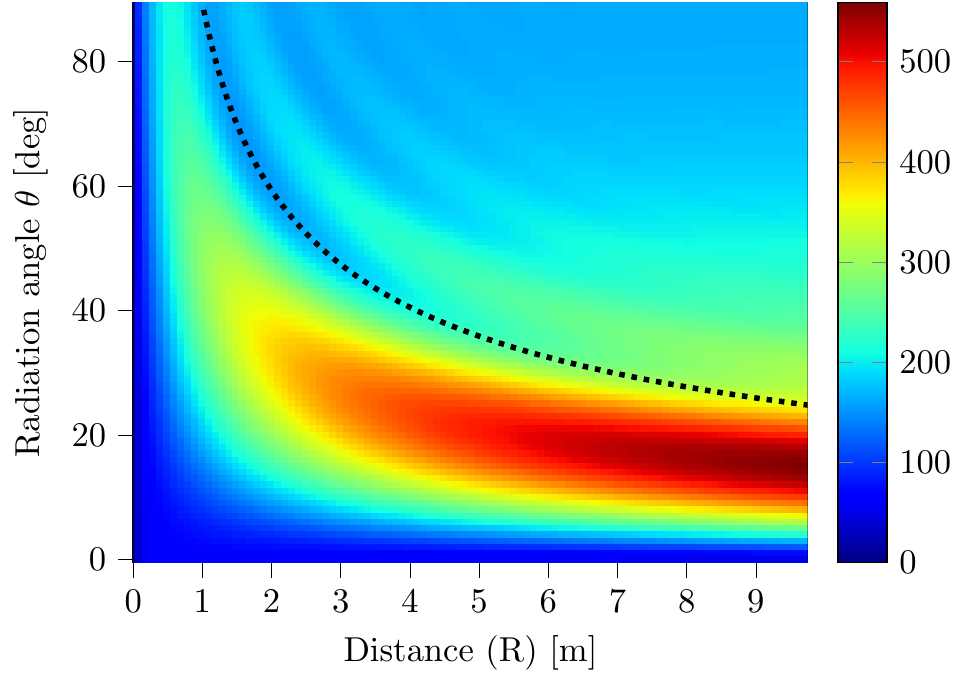}} 
\subfloat[]{\includegraphics[width=0.48\textwidth]{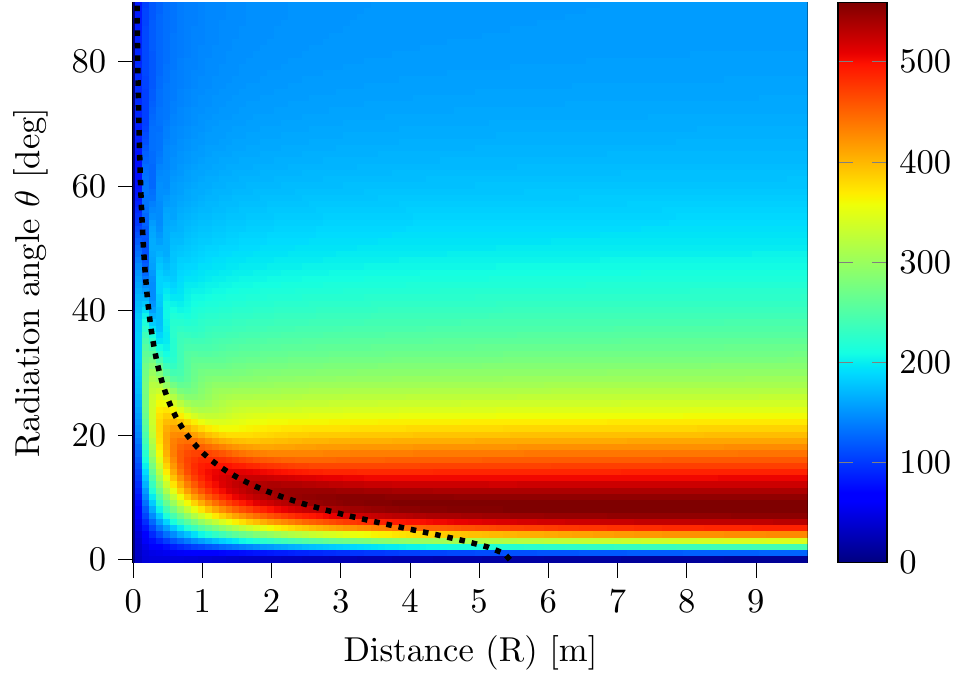}}
\caption{ The angular distribution 0.3 GHz and 5.7 GHz for $\beta = 0.99$ as a function of monitor distance from the target. The target size is set to 1.4$\beta\gamma\lambda$ for the reference wavelength of $\lambda_{ref}= 1$ m. Eq.~\ref{eq:formation_length} scaled with $2\pi$ is shown as dotted lines which separate wave and prewave zone.}
\label{fig:5}
\end{figure}

\begin{figure}[H]
\centering  
    \includegraphics[width=0.8\textwidth]{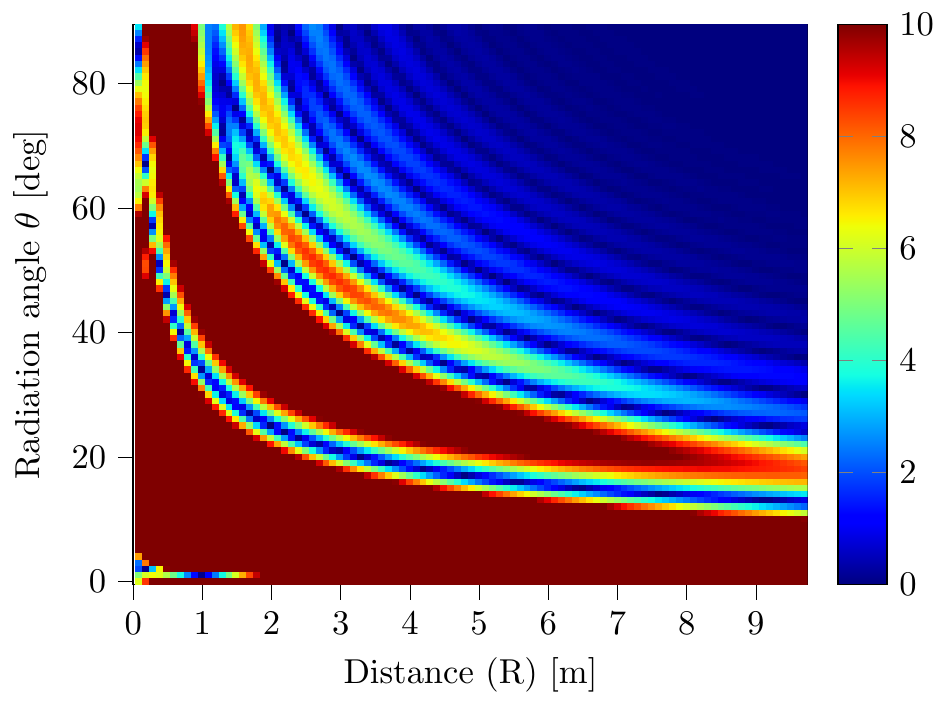}
    \caption{Percentage error between 0.3 GHz and 5.7 GHz for $\beta = 0.99$ as a function of monitor angle and distance from the target. The target size is set to 1.4$\beta\gamma\lambda$ for the reference wavelength of $\lambda_{ref}= 1$ m.  Above 2 m radial distance, the radiation has a far field like distribution for the relevant frequency range for angles larger than 60\degree.}
    \label{fig:6}
\end{figure}

Figure~\ref{fig:5} shows the angular field distribution for two wavelengths $\lambda = 1$ m and $\lambda = 0.055$ m for the upper end of our beam energy consideration i.e. $\beta=0.99$ ($\gamma = 7.08$) as a function of monitor distance and observation angle with respect to the TR target. Let us consider to frequencies 0.3 GHz and 5.7 GHz as the extremas for a 100 ps ($\sigma$) Gaussian bunch. The target size chosen is 1.4$\beta\gamma\lambda$ with the reference wavelength $\lambda_{ref} = 1$ m in order to discount the target size effects at frequencies above $300$ MHz. Note this leads to $b=\approx 20$ m target radius for $\beta=0.99$ case.  The Verzilov prewave zone formula predicts a $L_{formation}$ of $\approx 50$ m for $\lambda=1$ m and $\approx 2.5$ m for $\lambda=0.055$ m. One should note that Ginzburg "formation zone" is $2\pi$ times smaller and its angular dependence is marked in Fig.~\ref{fig:5}. For $\lambda=0.055$ m case, the far field angular field distribution is obtained (for all angles) from a monitor distance of 2.5 m (and onwards) in accordance to the estimate. It is also seen that the deviation from the far-field field distribution is a strong function of the angular placement of the monitor. Figure~\ref{fig:6} shows the relative error between the field angular distribution for the two frequencies. If we tolerate a 5\% relative error, already at $R > 2$ m and $\theta > 60$ degrees, the near field effects are negligible and the monitor for the full range of frequencies in an 100 ps bunch and the monitor can be considered in the wave zone. Thus, the field monitors can be placed much closer in comparison to the relativistic approximation $\gamma^2\lambda$ at steeper angles. This is a useful observation in context of measurement of longitudinal charge profiles since monitors can be placed relatively close to target for relevant energy range $\beta <0.99$ without relevant distortions of the near field effects. This also allows for higher signal intensity and flexibility in the set-up design.

\subsection {Effect of hole in the target}
\begin{figure}[H]
\centering 
\subfloat[]{\includegraphics[width=0.48\textwidth]{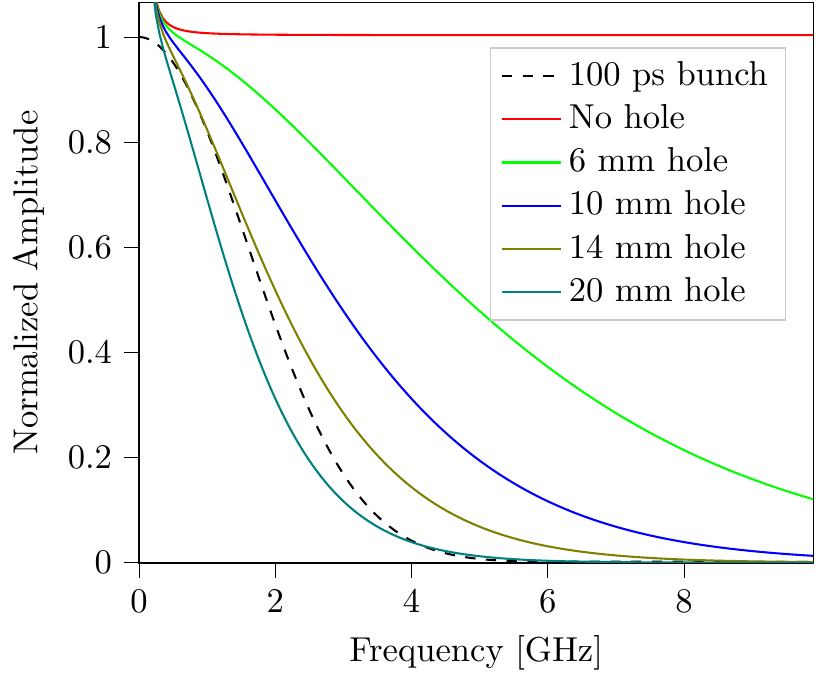}} 
\subfloat[]{\includegraphics[width=0.48\textwidth]{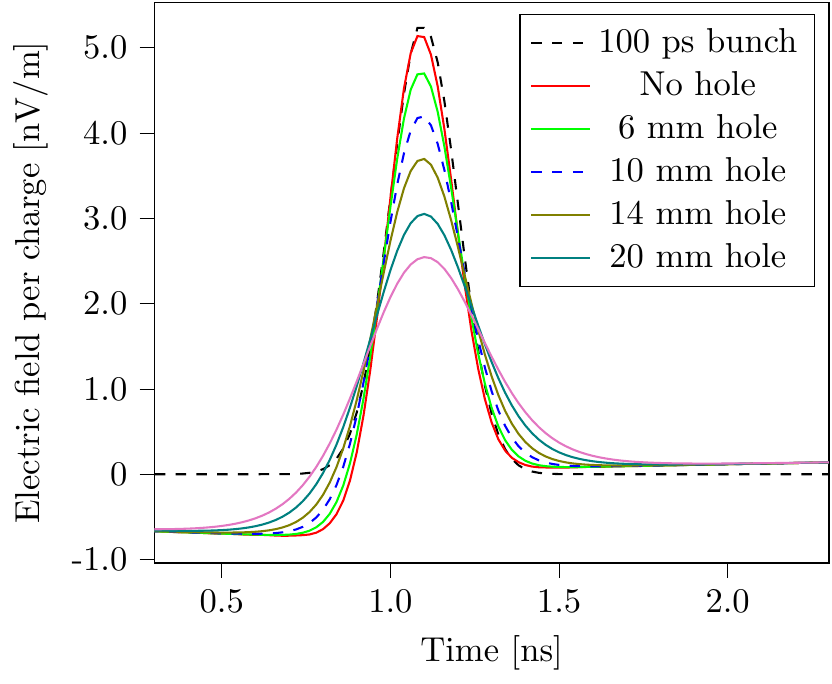}}
\caption{ (a) Frequency response as a function of hole size for $\beta = 0.15$ at monitor distance $R$ = 1.0 m. Frequency spectrum of a 100 ps is marked with dashed lines. (b) Measured electric field at 85 degrees for an input of 100 ps bunch for different hole sizes.}
\label{fig:7}
\end{figure}

Another important design consideration for any beam diagnostics device is its potential usage without interfering with the beam. Diffraction radiation is a closely related concept to transition radiation which can be utilized for non destructive measurements. Unlike transition radiation, the charges need not "hit" the second medium, but only pass close enough, such that the effective source size has significant interaction with the second medium. In our setup, a hole through the target provides such a possibility. A transversely small beam charge distribution will pass though the hole of diameter $d_{hole}$ ($2a$ in Fig.~\ref{fig:1}), however the 
effective size of the incident field $\beta\gamma\lambda$ will interact with the target to produce diffraction radiation. It is clear, for a wavelength $\lambda$ where $\beta\gamma\lambda < d_{hole}$, there is 
barely any radiation generated. The hole thus introduces a low pass behavior in the generation process. Figure~\ref{fig:7} (a) shows the frequency response calculated from the angular field distribution of $\beta = 0.15$ beam at monitor distance of $R = 1$m and $\theta = 85$ degrees and target size $\beta\gamma\lambda= 1$ for different hole sizes.
In addition, the frequency content of a 100 ps bunch is plotted for comparison. Figure~\ref{fig:7}(b) shows convolution of NIR frequency response with a 100 ps bunch for upto 10 mm holes are there is no significant widening of the bunch observed.
\subsection{Signal estimates}
An estimate of the expected induced voltages at a monitor placed at a distance of $R=1.0$ m and angle $\theta=$ 60 \degree for a  100 ps ($\sigma$) Gaussian bunch length containing 10 pC charge at the energy corresponding to $\beta = 0.15$ is sought in this section. Aforementioned charge per bunch coming at 36 MHz repetition corresponds to an average current of 0.37 mA. The target diameter should be larger than the effective source field, e.g. $d_{target} > 2\beta\gamma\lambda_{ref}$  for $\lambda_{ref} = 1$ m. This gives the lower bound on the target diameter of 0.3 m. 
The field estimate per charge per unit frequency can be obtained by inserting these parameters in Eq.~\ref{eq15}, which results in $\frac{dE_{TR}}{d\omega dq} = 4.0\cdot10^{-18}$ V/(m$\cdot$Hz). The field obtained per charge should be scaled with the number of charges $N=10pC/(1.6\cdot10^{-19}) = 6.25e7$.
For a bunch of charges with standard deviation $\sigma_t = 100ps$ of the longitudinal profile, most of the power lies below $2\sigma_f = \frac{1}{\pi\sigma_t} \approx 3.2$ GHz. The peak field estimate thus obtained is ~400 mV/m for 10pC charge in Gaussian 100 ps ($\sigma$) bunch. 
The antenna factor for a commercially available biconical antenna~\cite{Schwarzbeck} $\approx 30 dB$, i.e. $\frac{E_{field}}{V_{induced}}=30$ and the 12 mV peak voltage should be induced. As seen from Figure~\ref{fig:7}, a hole of diameter 10 mm will reduce the signal by a factor 2 and ~5 mV peak voltage is available for non destructive charge profile measurements. In comparison, a typical amplifier~\cite{AnalogDevices} with $50$ $ \Omega$ input impedance has a noise of $\approx 1$ $nV/\sqrt{Hz}$ which for a 10 GHz bandwidth should produce an rms noise of 0.1 mV. Thus a signal to noise ratio (SNR) of 50 is available for 10 pC charge in a 100 ps bunch measured with a commercially available biconical antenna place at $\theta = 60\degree$ at 1 m distance between target and detector. 

\subsection {Applicability range for beam and target parameters}
The longest bunches which can be measured by this method is limited by the target size which in turn is given by $\beta\gamma\lambda$ as well as the lower frequency cut-off of the commercially available linear phase broadband antennas. Both of the above conditions converge to a similar limit, i.e. a wavelength of $\lambda = 2$ m corresponding to a frequency of 150 MHz. This means that a bunch length of 500 ps (1 $\sigma$) could still be faithfully reconstructed. The shortest bunches which can be measured is primarily limited by the sampling speed and analog bandwidth of acquisition electronics and the current technology can allow 20 ps bunches to be measured. Since the transition radiation signal is proportional to beam velocity and current, beams upto $\beta = 0.01$ should be measurable provided there is enough beam current. Our range of application is $\beta = 0.05$ to $0.15$ at GSI UNILAC.

\section{CST simulations and comparison}

\begin{figure}[ht]
\centering 
{\includegraphics[width=0.68\textwidth]{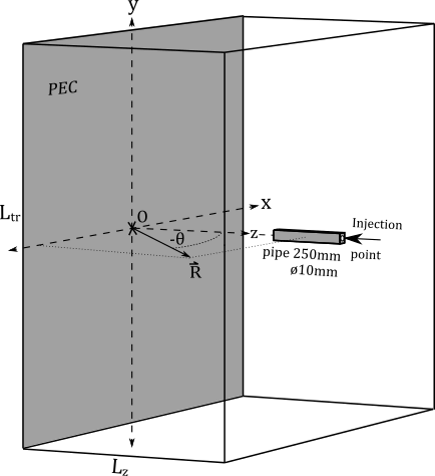}} 
\caption{The Simulation model}
    \label{fig:8}
\end{figure}

\begin{figure}[ht]
\centering 
\subfloat[]{\includegraphics[width=0.48\textwidth]{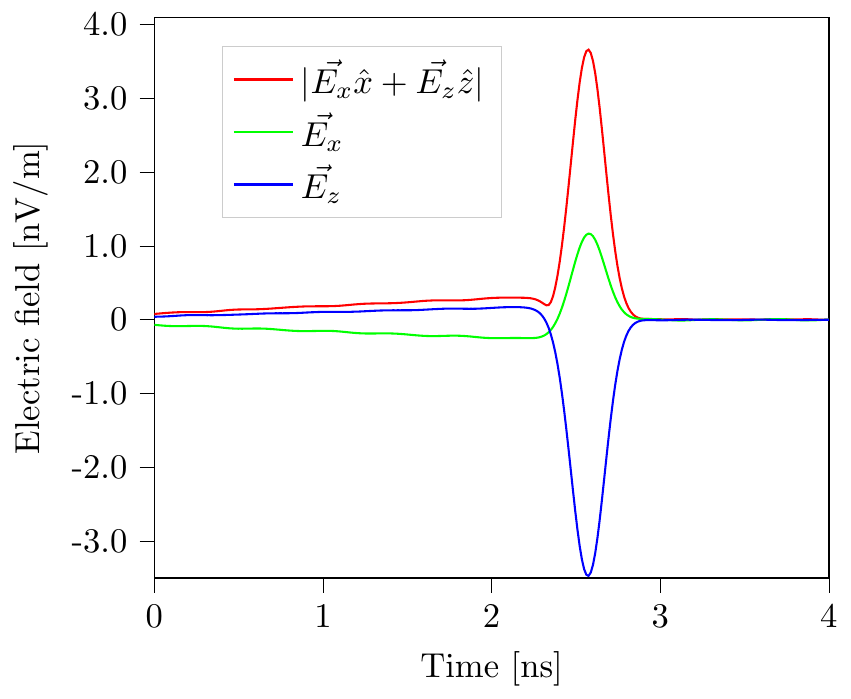}} 
\subfloat[]{\includegraphics[width=0.48\textwidth]{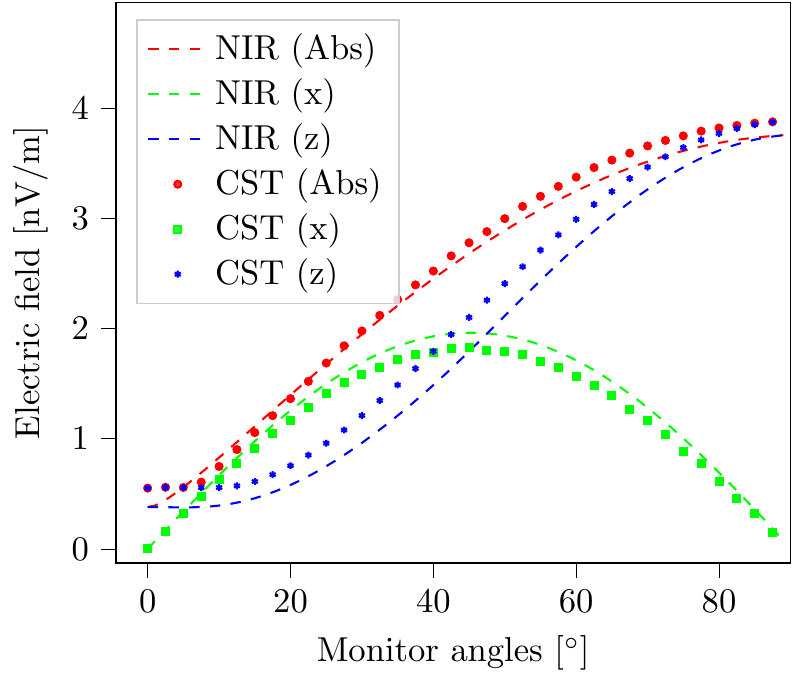}}
\caption{ (a)  Time domain snapshot of all three components of the transition radiation electric field for beam velocity $v = \beta \cdot c$ where $\beta = 0.15$ (b) Field amplitudes as a function of monitor angle obtained from CST in comparison with NIR. }
    \label{fig:9}
\end{figure}

Transition radiation process was simulated using the particle-in-cell (PIC) solver of the electro-magnetic simulations software CST. Here a collection of macro-particles with common mass $m_q$, total charge $q$ and velocity $\beta$ form a bunch with user-defined transversal as well as longitudinal shape. Figure~\ref{fig:8} shows the simulation domain. The transverse beam size is fixed to 5 mm $(\sigma)$. This charged particle bunch is introduced into and propagated through a rectangular calculation domain (here: vacuum) which is spatially discretized by a regular mesh. The temporal discretization complies to the Courant–Friedrichs–Lewy criterion~\cite{Courant} and is provided by the solver algorithm. After propagation through the calculation domain, the bunch is normally incident on a target made of a perfect electric conductor (PEC) spanning fully over one of the boundary planes. Optionally it has a hole in the center to let the bunch particles pass through. Here the wanted transition or diffraction radiation is formed since the self-field of the bunch has to comply dynamically to the PEC interface as discussed earlier in this report. All remaining boundaries are set to be \textit{open} so that time-dependent electro-magnetic fields are absorbed with a low level of residual reflections which is realized by imposing perfectly matched layers (PML). 
A drawback of the simulation via the PIC solver is given by the fact that the insertion of a charge into the simulation domain results in prompt bremsstrahlung which we refer to as "domain entry radiation" for the lack of a better term. It is briefly discussed in the next section.
The electric fields produced during the simulation are collected with ideally broadband field monitors radially ($R$) and azimuthally $\theta$ around the impact point. 
We choose a Gaussian longitudinal distribution of the bunch charge with tails cut at $4~\sigma$ and uniform velocity $\beta \in [0.15, 0.99]$. The range of bunch lengths is given as $\sigma_t \in [100, 400]$~ps thus a conservative upper cutoff frequency of $6.5$~GHz is used to contain the frequency span of the shortest bunch. Behind the entry pipe, an electrically large problem is faced here and the mesh cell size is controlled via the spatial sampling of the smallest wavelength. The relevant parameter \textit{lines per wavelength} (lpw) is set to at least lpw $= 11$ to avoid numerical dispersion. 
To ignore any potential space charge induced bunch blow-up the specific charge of the macro-particles is set to (unphysical) $10^{-4}~$C/Kg. In accordance with the simulated velocity and thus the effective source size the domain size has been varied between 0.5~m and 2~m in transversely and between 1~m and 3~m in longitudinally marked as $L_{tr}$ and $L_z$ in Fig.~\ref{fig:8}.
Figure~\ref{fig:9}(a) shows the output of the field monitors placed at $\theta=70\degree$ and distance of $R=1$ m. The beam velocity corresponds to $\beta = 0.15$ and the bunch length is set to 100 ps $\sigma$.  The domain parameters are $L_{tr} = 1.25$ m and $L_z = 1.0$ m .$\vec{E_x}$, $\vec{E_z}$ as well as absolute field value is shown.  Figure ~\ref{fig:9}(b) shows a comparison of the peak fields seen by a perfect broadband monitor at 1 m distance obtained by CST simulations from 0 to 90\degree  in comparison to angular distribution from the NIR (Eq.~\ref{eq15}). For NIR, angular distributions of radiated electric field for 100 ps bunch is calculated as follows,
\begin{align}\label{eq:NIRweighting}
    Ef(\theta, R)= \frac{\sum_{f=1}^{N} W_f \cdot NIR(\theta,R,f)}{\sum_{f=1}^{N}W_f}
\end{align}
where f depicts the frequency from 0.05 to 6 GHz with a resolution of 0.05 GHz. $NIR(\theta,R,f)$ can be obtained from Eq.~\ref{eq15} for a given angle, distance and frequency.
The weights $W_f$ are according to relative amplitudes at the given frequency for $\sigma = 100$ ps bunch, i.e. if $g(t) = \exp{-\frac{t^2}{2\sigma^2}}$,
and $W_f = DFT[g(t)]$
The absolute estimates obtained for Eq.~\ref{eq:NIRweighting} is 30\% lower than the peak values obtained in CST. The peak values obtained in CST are scaled before plotting for easier comparison of relative strengths. However the relative agreement between the angular distributions obtained via CST and NIR is rather good.

\begin{figure}[H]
\centering 
\subfloat[]{\includegraphics[width=0.48\textwidth]{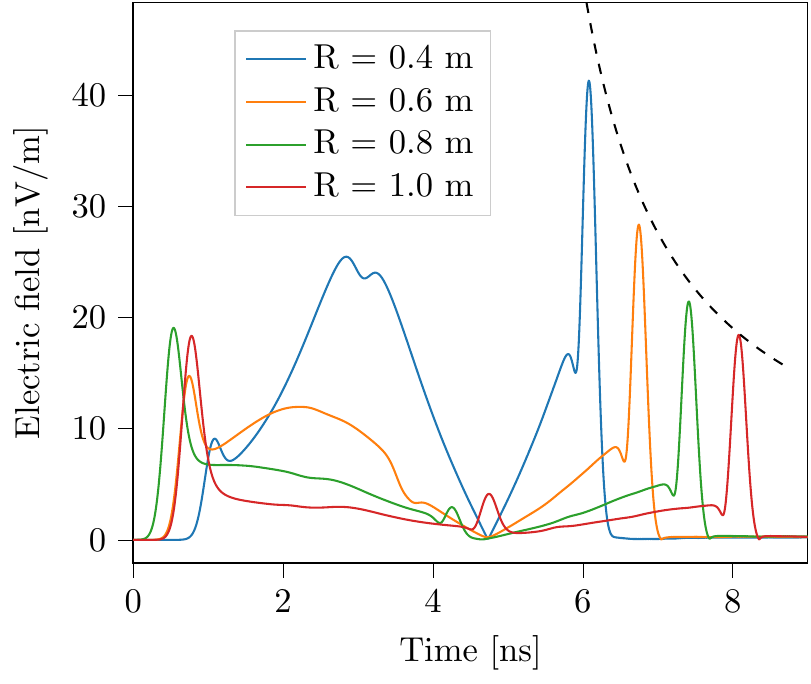}} 
\subfloat[]{\includegraphics[width=0.48\textwidth]{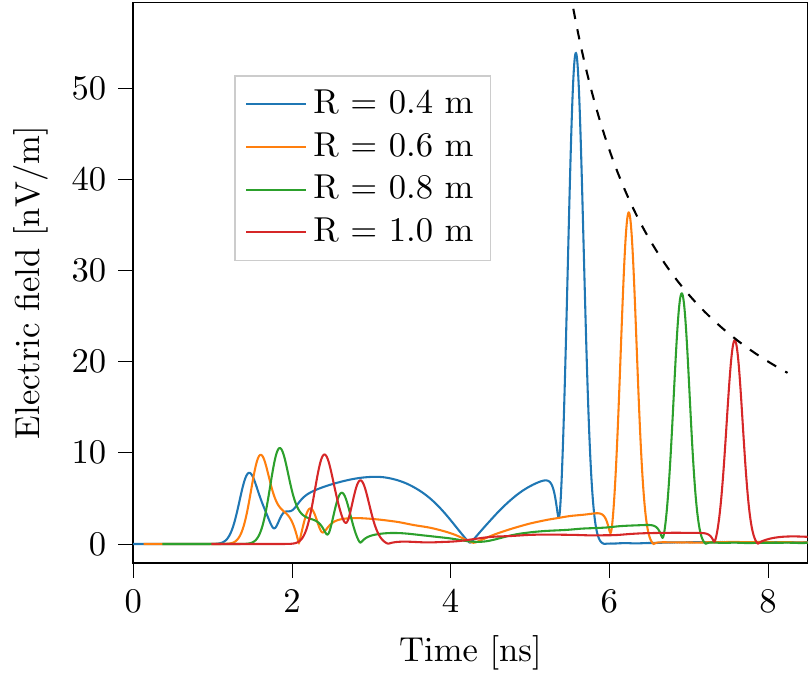}}
\caption{Absolute TR electric fields at 4 monitors placed at incremental distances from the target at (a) 45\degree and (b) 80\degree angles for $\beta = 0.6$.}
\label{fig:10}
\end{figure}

Figure~\ref{fig:10} shows the absolute electric fields at 4 monitors placed at incremental distances from the target at 45\degree and 80\degree angles for $\beta = 0.6$. The first peak at 1 ns is the diffraction radiation generated when the charge exits the pipe. Following that, direct fields of the beam are seen until the beam crosses the target at 4.5 ns when the transition radiation is generated and detected at various monitors placed at incremental distances. The direct field is stronger for monitors placed at 45\degree due to proximity to the traversing charge. This proximity effect and direct field estimates are shown in the appendix. It should be noted that at 85\degree the peak heights at different distances closely follow the $1/r$ dependence while at 45\degree there is a slight deviation. This is in line to the discussion in earlier sections concerning the dependence of formation zone on detection angle $\theta$.
\subsection {"Entry" radiation}
Unlike in the most theoretical approaches where the charge usually comes from infinity onto the target, there is a the production of "entry" radiation when charge enters computational domain in our PIC simulations. This is thus seen as an artifact and complicates comparison between simulation and theory. The production of this kind of bremsstrahlung on domain entry in terms of amplitude and phase is expected to be dependent on the combination of mesh cell size and velocity which was experienced in our preliminary simulations. To prevent the propagation of the radiation generated at charge entry, the bunch is introduced into the domain in a small pipe of length $250$~mm as shown in Fig.~\ref{fig:8}. The bunch exiting the pipe produces diffraction radiation which is still an unwanted artifact yet more stable w.r.t. mesh cell size and beam velocity. The choice of the pipe diameter is a tradeoff, it should be small such that the excitable waveguide modes are well above the set frequency cutoff of the simulation, but if its too small, significant diffraction radiation is generated which interferes with the transition radiation for small angles $\theta$ with respect to beam axis. We use the pipe diameter to 30 mm in most cases except for one case where 100 mm was chosen to suppress diffraction radiation.  

\subsection{Angular distribution at different beam velocities and comparison with analytical estimates}
\begin{figure}[ht]
    \centering
    \includegraphics*[width=8.5cm]{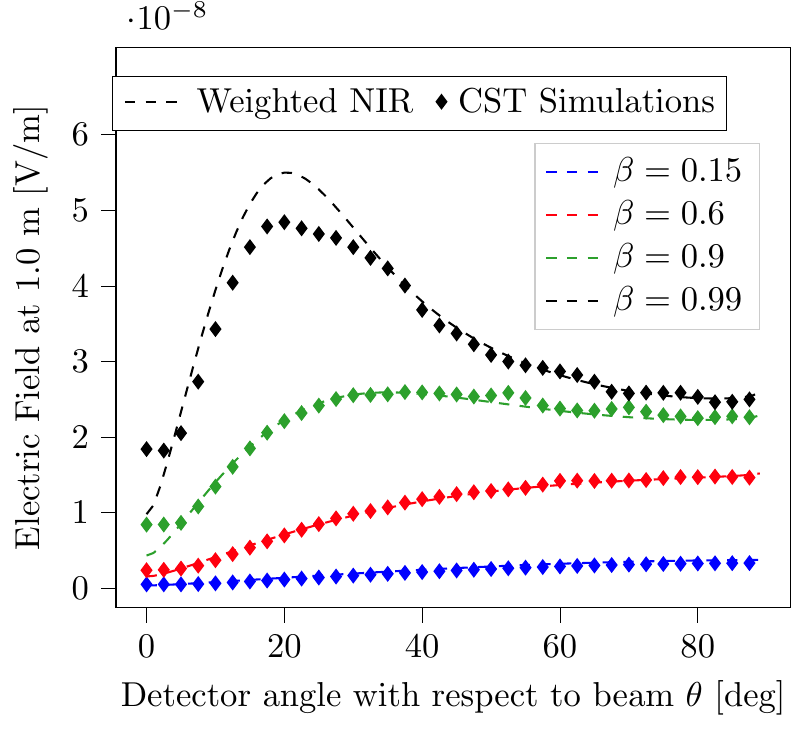}
    \caption{Comparison of analytical and simulation angular field distribution for different betas at $R= 1.0 m$ distance.}
    \label{fig:11}
\end{figure}

Figure~\ref{fig:11} shows the angular distribution of absolute value of E-fields calculated using NIR for 100 ps bunch in the same way as discussed in previous section (Eq.~\ref{eq:NIRweighting} and the peak field recorded on the CST broadband monitors. 
The dimensions of the domain were made longitudinally large and transversally asymmetric in order to have a reasonable calculation time, i.e. $L_z = 3.2$ m, while $L_{tr,x}= 2.0$ m and $L_{tr,y}= 0.6$ m. As we know from the first section, the TR detected in horizontal plane is produced by the corresponding vector currents in the same plane and the orthogonal components make no contribution to each other. The entry pipe radius used in this case was chosen to be of 100 mm diameter because the diffraction radiation at the exit of a smaller pipe for higher betas $\beta=0.9,0.99$ was significantly distorting the angular distribution at smaller angles. The disadvantage of larger pipes is that waveguide modes (here predominantely $TE_{11}$ with $f_{cut} \approx 1.76$ GHz) are excited in the pipe and make it to the monitors which are seen as wiggles in the angular distribution.

\begin{figure}[H]
\centering 
\subfloat[]{\includegraphics[width=0.48\textwidth]{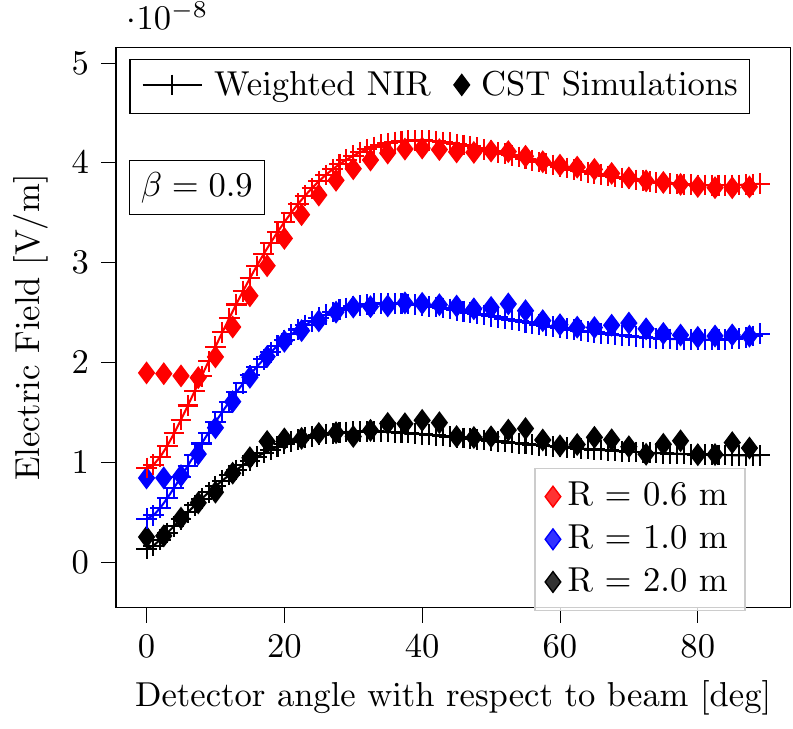}} 
\subfloat[]{\includegraphics[width=0.48\textwidth]{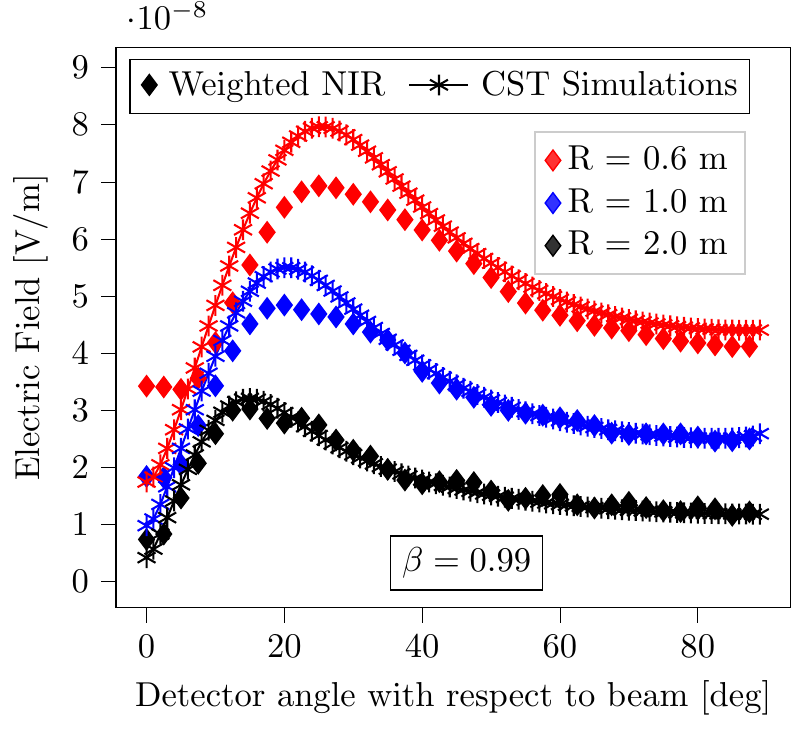}}
\caption{ The angular distribution 0.3 GHz and 5.7 GHz for $\beta = 0.99$ as a function of monitor distance from the target. The target size is set to 1.4$\beta\gamma\lambda$ for the wavelength of $\lambda= 1 m$.}
\label{fig:12}
\end{figure}
Figure~\ref{fig:12} shows the comparison of the angular distribution for the NIR and CST monitors at three distances for  $\beta=0.9,0.99$ and the dependence is similar as a function of monitor distance. The deviation for $\beta=0.99$ for smaller angles $\theta < 40 \degree$ is primarily due to the influence of "entry" radiation.

\subsection{Effect of transverse target and hole size on radiation pattern and field intensity}
\begin{figure}[H]
\centering 
\subfloat[]{\includegraphics[width=0.48\textwidth]{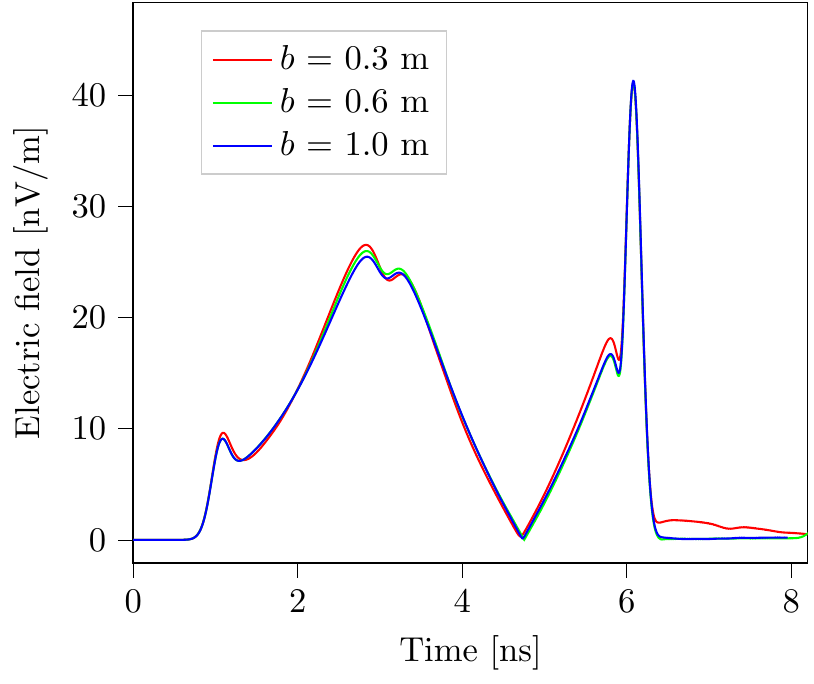}} 
\subfloat[]{\includegraphics[width=0.48\textwidth]{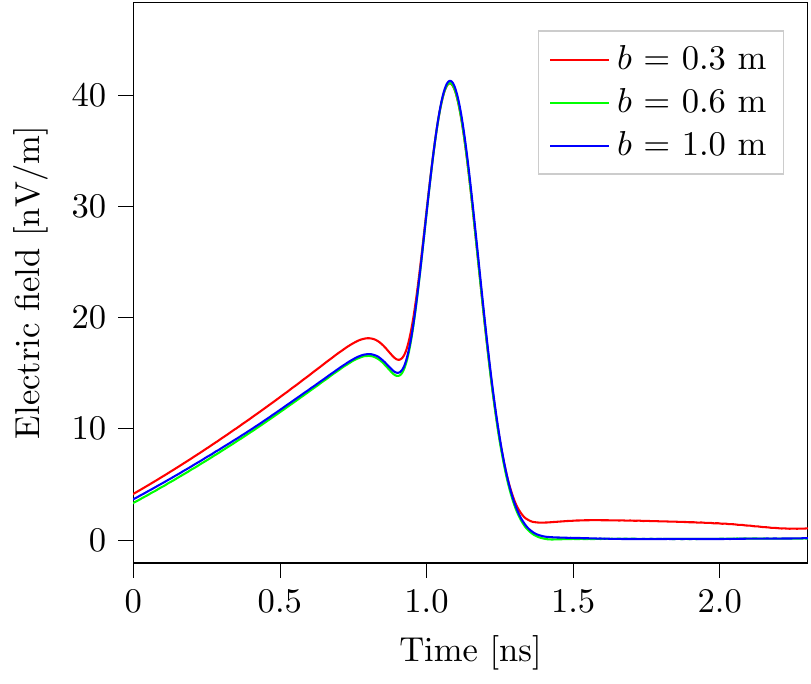}}
\caption{ (a) Obtained radiation (Absolute values of E field) as a function of target size for $\beta = 0.6$ at monitor distance $R$ = 0.4 m at 45 degrees.  (b) Selected region to highlight the differences.}
\label{fig:13}
\end{figure}
The transverse target size was varied from 0.3 m to 1.0 m (half width) for the beam velocity of $\beta=0.6$. The target size was varied along with the transverse domain size in order for the target to touch the open boundaries and avoid the possibility of build up of a floating potential on the target. The absolute value of field measured at a monitor placed and 40\degree and 0.4 m is shown in Fig.~\ref{fig:13} (a). Figure~\ref{fig:13} (b) shows the radiation peak without the entry radiation and the direct field parts for better visibility of the radiation difference due to target size differences. Minor differences in the shape of the measured bunch are seen, and it appears that some low frequencies cut off for target size $b = 0.3$ m which is inline with the expectation. 

\begin{figure}[H]
\centering 
\subfloat[]{\includegraphics[width=0.48\textwidth]{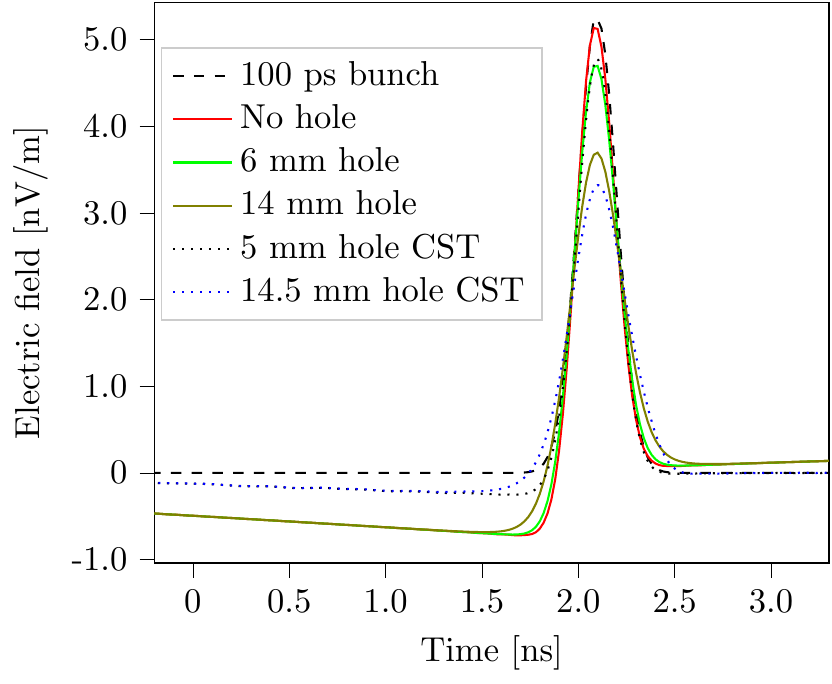}} 
\subfloat[]{\includegraphics[width=0.48\textwidth]{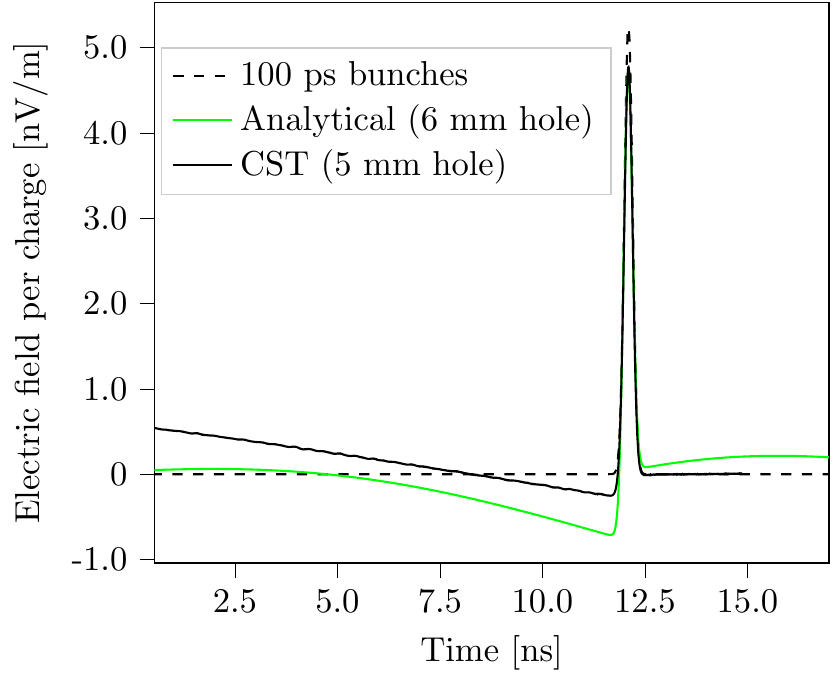}}
\caption{ (a) Obtained radiation (z component of E field) as a function of hole size for $\beta = 0.15$ at monitor distance $R$ = 1.0 m at 85 degrees for an input of 100 ps $\sigma$ bunch for different hole sizes.(b) Zoomed out view of the plot on left.}
\label{fig:14}
\end{figure}
Similarly the size of the hole in the center of the target for the non-intrusive passage of the beam is varied. Figure~\ref{fig:14}(a) shows the $z$ component of the electric field observed at R= 1.0 m at 85 \degree for a 100 ps $(\sigma)$ bunch.  The no-hole case is compared with the NIR estimate of 6 mm and 14 mm hole (Same as presented in fig.~\ref{fig:7}) as well as CST calculation of 5 mm and 14.5 mm hole radius. NIR estimated temporal profile is obtained by convolving a 100 ps bunch with the frequency response at the given $\theta$ and $R$. There is a good agreement between NIR and CST calculations and the charge distribution is faithfully reproduced even for the largest hole size chosen.

\section{First prototype and results}

The GSI UNILAC facility provides wide range of ions from 1.4 MeV/u to 11.4 MeV/u. The expected bunch lengths range from 100-500 ps ($\sigma$) with a repetition rate of 36 MHz or 108 MHz depending on the injector used.  The currents are in the order of 50 uA to few mA. 
\begin{figure}[ht]
    \centering
 \subfloat[]{\includegraphics[width=0.43\textwidth]{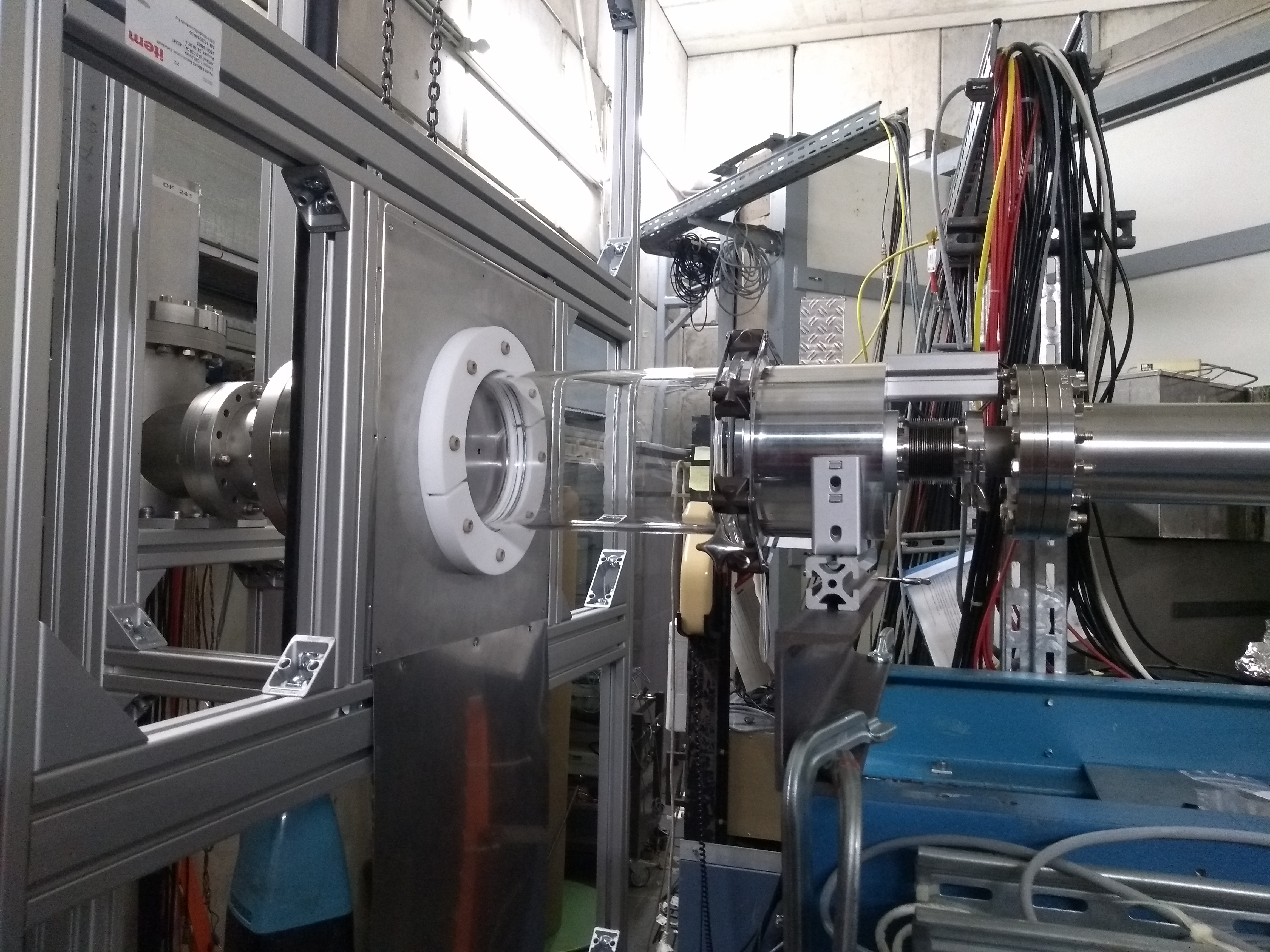}}
 \subfloat[]{\includegraphics[width=0.54\textwidth]{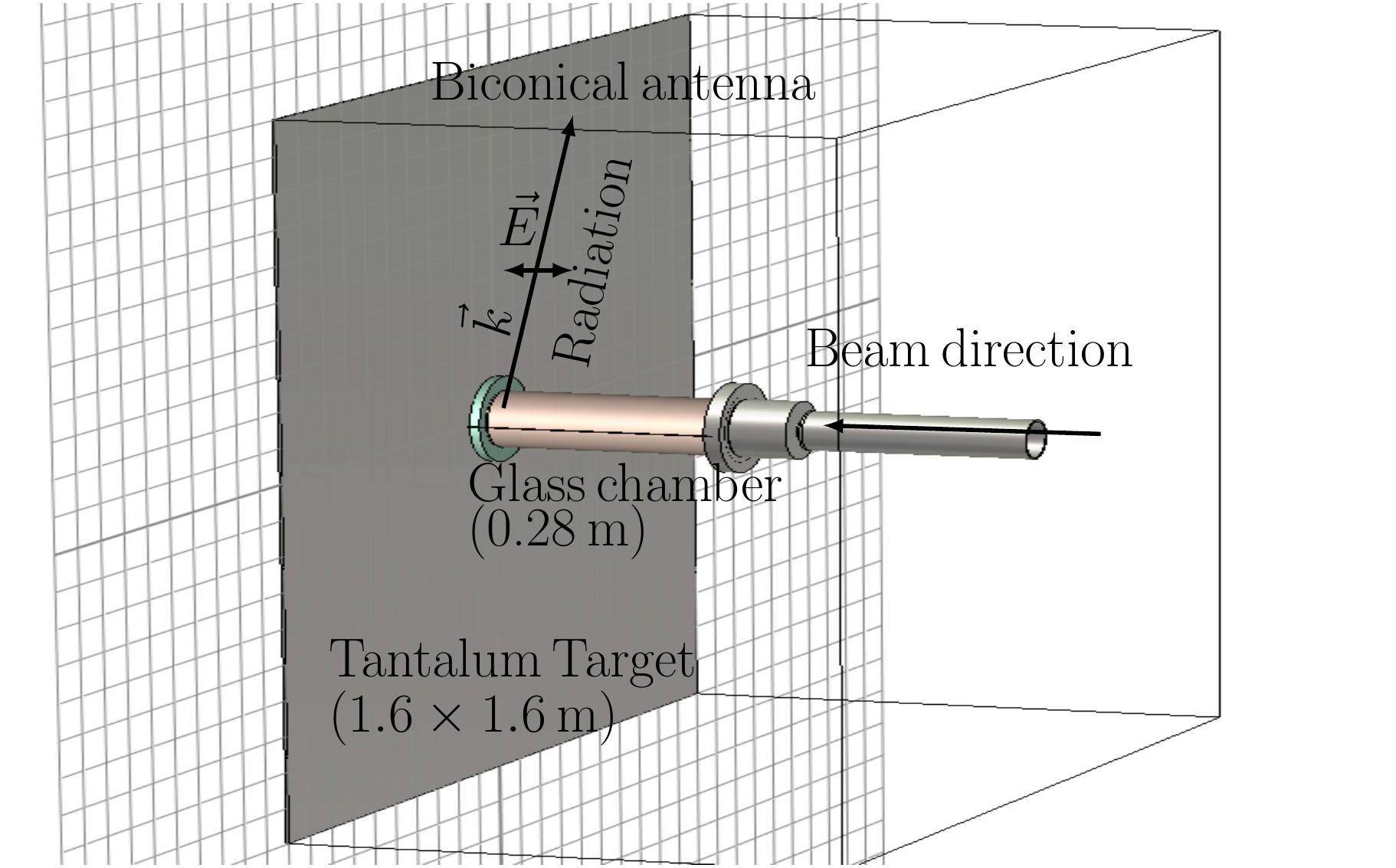}}
    \caption{(a) The Simulation model of the set-up. (b) Photo of the first prototype set-up.}
    \label{fig:15}
\end{figure}
The main consideration of the TR set-up design is the dimensions and material of the RF window, transition section between the beam pipe and the glass, as well as target size and hole size. On the detector side, the Antenna gain and phase response are important to evaluate.

\subsection{Set-up details and simulations}
\begin{figure}[H]
\centering 
\subfloat[]{\includegraphics[width=0.48\textwidth]{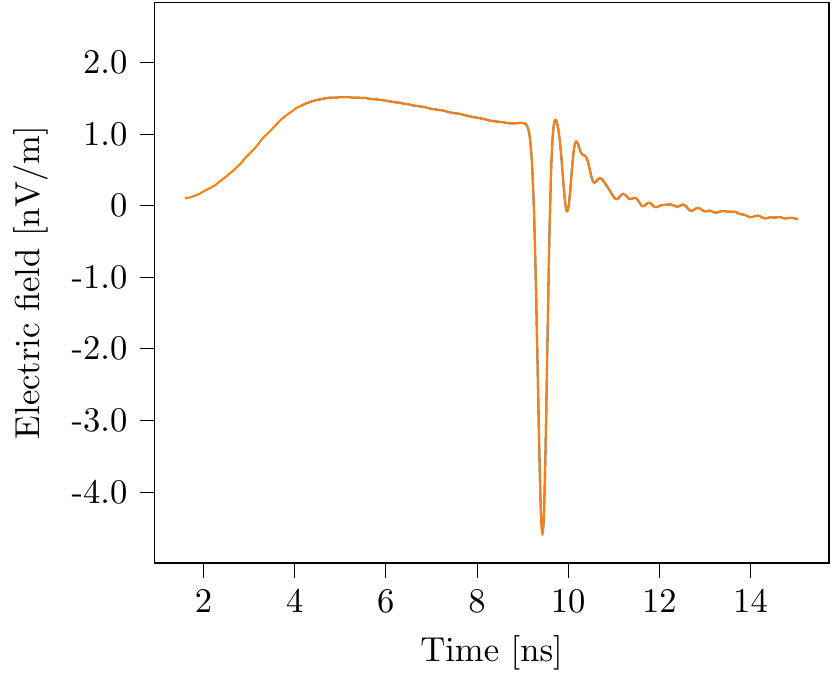}} 
\subfloat[]{\includegraphics[width=0.48\textwidth]{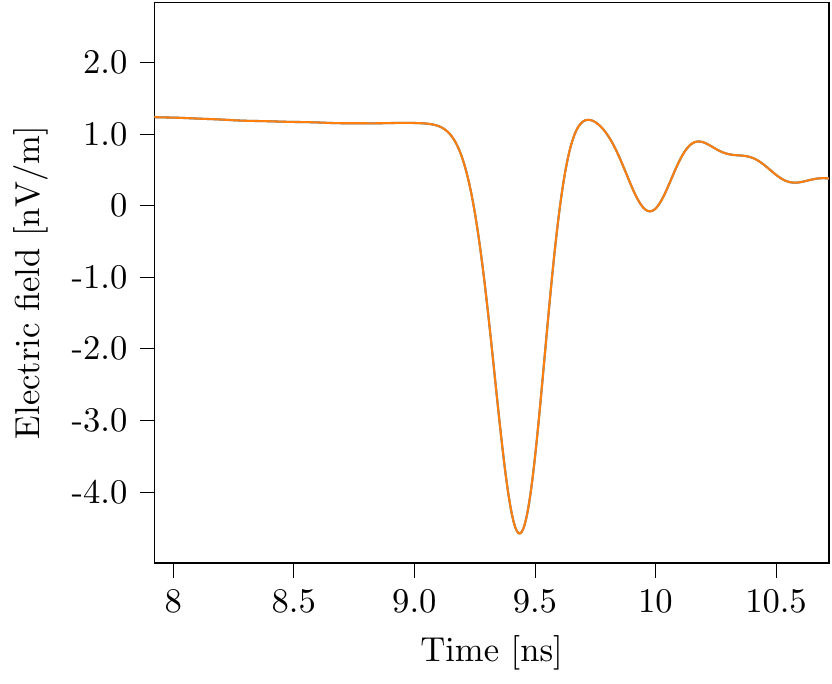}}
\caption{ (a) CST simulation output showing $\vec{E_z}$ component at the broadband monitor of the model shown in Fig.~\ref{fig:15} (b) for 100 ps Gaussian charge distribution traversing with velocity $\beta=0.15$. (b) Enlarged section around the bunch signal.}
\label{fig:16}
\end{figure}

   Figure~\ref{fig:15}(a) shows the photo of the first prototype of the set-up at the end of X2 experiment cave beamline at GSI. The last quadrupole was about 3m upstream, current measurement device was 1 m upstream and profile grids 1.5 m upstream. We chose the largest fused silica (glass) available off-the-shelf as the RF window with some custom connectors for a vacuum tight assembly with the beam pipe and the TR target. A Tantalum target (GTR target) with 3 mm hole in the center was used as target plate. A sensitive Faraday cup was installed just behind the hole to ensure that the beam was hitting the target since the alignment of the beam was tricky due to a small flange between beam-pipe and RF window.
   Figure~\ref{fig:15}(b) shows CST model of the proposed schematic for a set-up. Fig.~\ref{fig:16} (a) shows the $\vec{E_z}$ component detected at the monitor for a 100 ps bunch, with $\beta=0.15$ at 1.0 m distance. Direct field is observed between 2 and 9 ns ending with the TR field peaking at 9.4 ns. The enlarged plot around the TR radiation in Fig.~\ref{fig:16}(b) and the TR reflection from the glass chamber is also seen at 10 ns.
The detector for the generated pulse was a Biconical antenna~\cite{Schwarzbeck}. Relevant characteristics of the antenna were measured as shown in the Appendix. A ring pick-up similar to shown in Figure 5.7 in ~\cite{Forck} is installed ~2m upstream of the GTR target which was used to trigger the digitizer. The pick-up signal and the Antenna output were connected to the same 80 GSa/s digitizer with an analog bandwidth of 20 GHz. Full macropulse of 100 $\mu$s signal was recorded for many successive pulses. 
\subsection{Pick-up vs GTR measurement}
\begin{figure}[H]
\centering 
\subfloat[]{\includegraphics[width=0.48\textwidth]{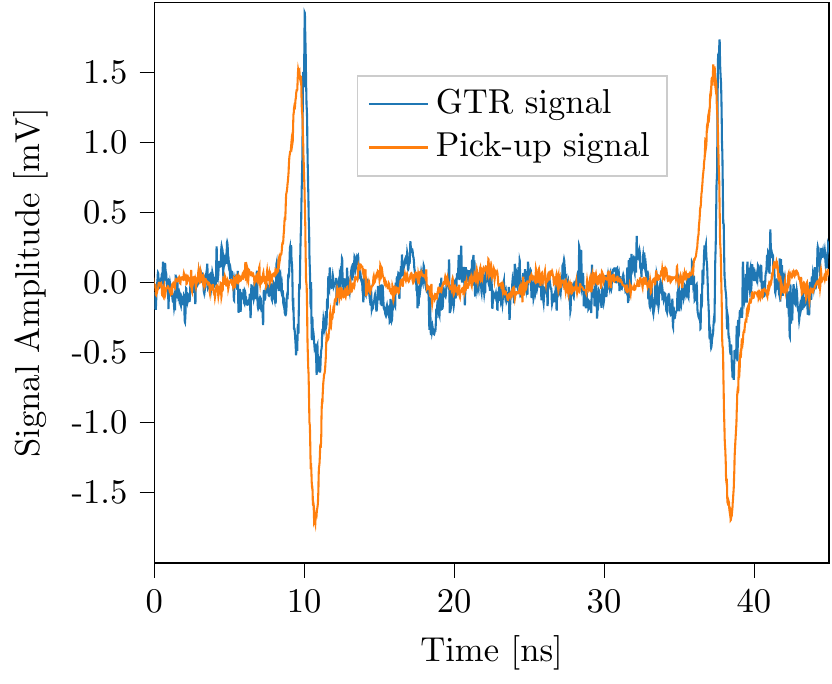}} 
\subfloat[]{\includegraphics[width=0.48\textwidth]{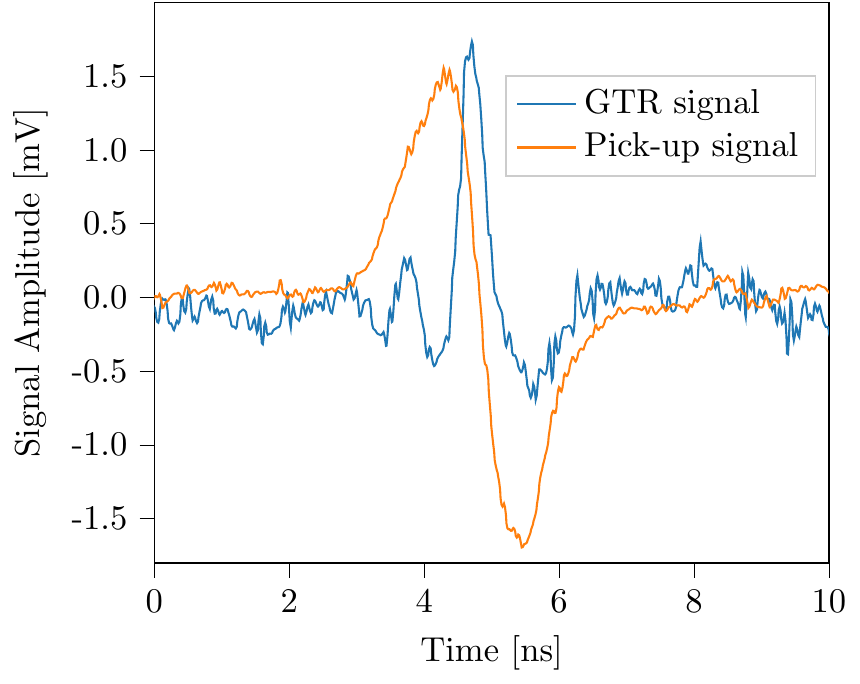}}
\caption{ (a) Two consecutive bunches measured by a pick-up and GHz transition radiation monitor spaced apart by ~2 m (b) Enlarged view focussing on the second bunch.}
\label{fig:17}
\end{figure}

Figure~\ref{fig:17}(a) shows two consecutive single shot bunches measured by the pick-up and the GTR antenna. The pick-up signal is divided by 10 for fitting on the same y-axis and time (x-axis) is translated by 42.5 ns to account for the particle traversal from pick-up to GTR target along with the signal propagation from GTR target to the antenna. Figure~\ref{fig:17}(b) shows the zoomed view of the second bunch and an additional structure on the pulse which is clearly visible on the GTR signal which is also hinted on pick up signal (but not resolved). The signals measured are about factor 5 smaller than expected from analytical estimates. Our suspicion in hindsight is that; most of the beam was lost at the boundary of beam pipe and RF window.

\subsection{Shot-to-shot fluctuations in charge distribution}
\begin{figure}[ht]
    \centering
 {\includegraphics[width=0.95\textwidth]{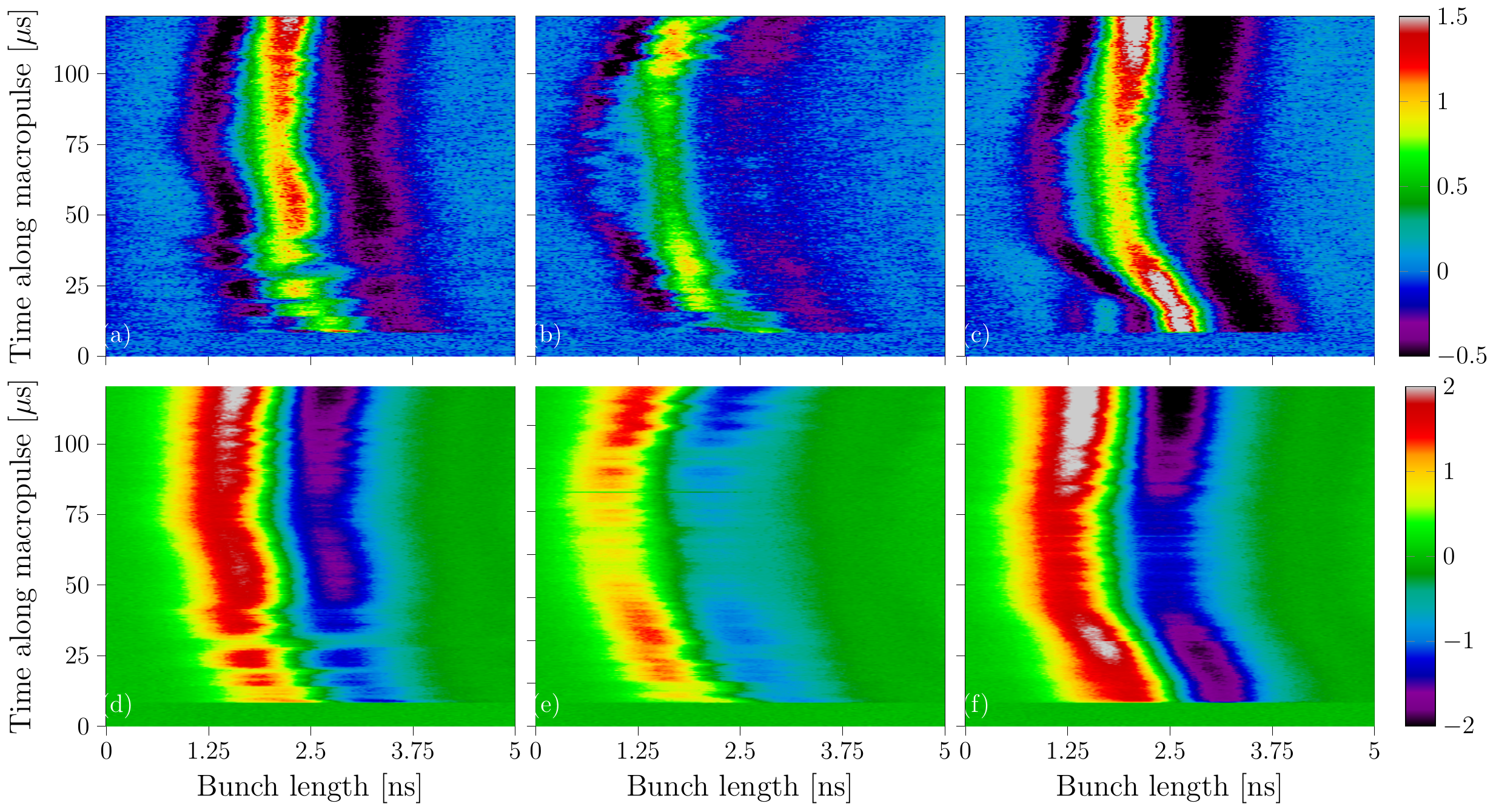}}
    \caption{Bunch length measurement for three consecutive macropulses by GTR (top row) and Pick-up (bottom row).}
    \label{fig:GTRBPM}
\end{figure}
Figure~\ref{fig:GTRBPM} shows three consecutive macropulse images from GTR (top row) and BPM (bottom row). The x-axis is a region selected within the rf period where the pulses are present while on the y-axis the bunch length evolution along the macropulse is shown. There seems to be slight energy shift as well as significant shot-to-shot bunch shape variation. This is currently attributed to the beam loading in the LINAC and the detailed and precise cause is under investigation and outside the scope of this report. These measurements also highlight that an averaged measurement for such pulses could result into spurious measurements and bunch-by-bunch shot-by-shot non-destructive bunch length measurement could be important for high intensity LINAC optimization.
\section{Summary}
We have presented a concept for time domain monitoring for longitudinal charge profiles for non relativistic beams based on coherent transition and diffraction radiation in GHz regime. We have extended the quasi spherical approximation presented in~\cite{Fiorito} to an exact expression in near field for normally incident beam. Signal estimates for typical beam intensities using commercially available detectors are  provided. The derived near field expressions were compared with CST simulations and a good agreement is found. Finally, first prototype measurements were performed which confirm the signal estimates and proves applicability of this method for non destructive longitudinal charge profile monitoring. This method forms a building block for a compact 6D phase space measurement system when combined with non-destructive transform profile monitoring using transition radiation in optical regime~\cite{Singh_OTR}. 
Although we have focussed on non relativistic bunch charge distribution in this case, variants of this technique could also be helpful in storage rings for very small bunches i.e.  where the pick-up bandwidth pose limitations.
\section{Acknowledgment}
B. Walasek-Hoehne is gratefully acknowledged for the support and discussions. P. Forck and C. Krueger are acknowledged for providing support in data acquistion part. M. Mueller, A. Abou Abed are acknowledged for help with the GTR mechanical set-up.
\section{Appendix}

\subsection {Direct field estimates on the monitor}

\begin{figure}[ht]
\centering
\begin{tikzpicture}

\draw[ultra thick] (1,5)--++(270:5.0);
\draw[ultra thick,latex'-,red,dotted] (1,5)++(270:5.0/2)--++(180:3.5);
\draw[thick,-latex',dashed] (1,5)++(270:5.0/2)--++(150:3.2);
\draw[thick,-latex',dashed] (-1.8,2.5)--++(90:1.5);

\newcommand\shadeang{30}
\newcommand\shadelen{0.5}
\foreach[count=\i, evaluate=\i as \x using (\i*0.5)] \y in {0,1,...,10}
{
\draw[thick] (1,5)--++(270:\x-0.5)--++(\shadeang:\shadelen);
}
\draw[thick,dotted, ->] (3,4.0)--++(90:1.0);
\draw[thick,dotted, ->] (3,4.0)--++(0:1.0);
\draw[ultra thick, ->] (0.0,2.5) arc (160:130:1.0);

\node[thick,text width=1.0cm,align=center] (Medium1) at (-2.0,1.0) {Vacuum};
\node[thick,text width=2.0cm,align=center] (Plate) at (2.5,1.0) {Conductor};
\node[thick,text width=4.0cm,align=center] (bc) at (-1.0,2) {$\vec{\beta}c$};
\node[thick,text width=4.0cm,align=center] (bc) at (-1.8,2) {$q$};
\node[thick,text width=4.0cm,align=center] (r) at (-1.0,3.9) {$\vec{r}$};
\node[thick,text width=4.0cm,align=center] (R) at (-2.1,3.2) {$\vec{R}$};
\node[thick,text width=4.0cm,align=center] (M) at (-2.0,4.4) {$M$};
\node[draw,thick,circle,text width=0.1cm,fill,align=center] (Q) at (-1.8,2.5) {};
\node[draw,thick,circle,text width=0.05cm,fill,align=center,blue] (P) at (-1.8,4.0) {};
\node[thick,text width=4.0cm,align=center] (theta) at (-0.4,2.7) {$\theta$};
\node[thick,text width=4.0cm,align=center] (x) at (2.8,4.8) {$\hat{x}$};
\node[thick,text width=4.0cm,align=center] (z) at (3.5,3.8) {$\hat{z}$};
\end{tikzpicture}
\caption{Schematic of a charge passing by a monitor towards a target where it is normally incident.}
\label{fig:directfield}
\end{figure}
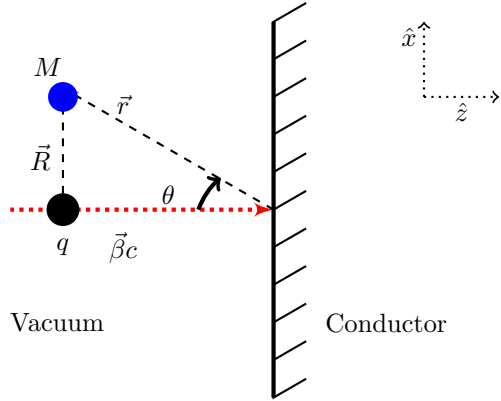
The direct field of the charge travelling at a velocity $\beta$ as shown in Fig.~\ref{fig:directfield} is given as\cite{JDJ},

\begin{align}
E_{direct} = \frac{q}{4\pi\epsilon_0}\cdot\frac{\gamma R\hat{e}_x+\gamma\beta c t\hat{e}_z}{\Big [R^2+ (\gamma\beta c t)^2\Big ]^{3/2}} 
\end{align}

where $t = 0$ when the charge has smallest Euclidean distance to the monitor $R$. The peaks for $E_{direct,x},E_{direct,z}$ occurs at $t=0, \pm R/\sqrt{2}\gamma\beta c$ respectively.

\begin{figure}[ht]
    \centering
 {\includegraphics[width=0.75\textwidth]{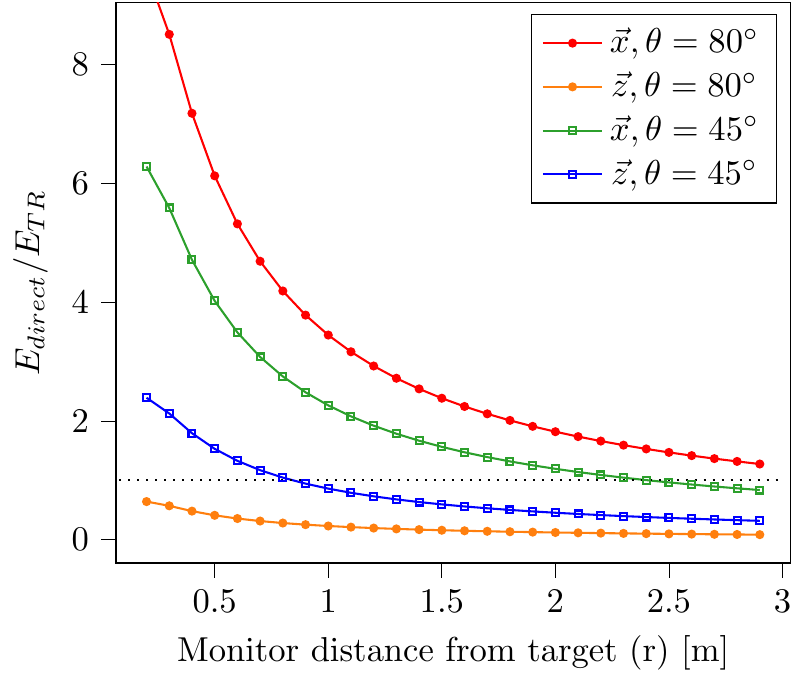}}
    \caption{Ratio of direct field against transition radiation field as a function of monitor distance from target for two monitor angles.}
    \label{fig:DirectvsTR}
\end{figure}

The charge is incident normally at the target and monitor distance is $r$ as shown in Fig.~\ref{fig:directfield}, while $R=r\sin{\theta}$. The transition radiation field per unit frequency  originating at the target is given by Eq.~\ref{eq15}. In the far field, Eq. 15 reduces to a simpler result~\cite{Garibyan}.

\begin{align}
\frac{dE_{TR}}{d\omega} = \frac{q\beta}{2\pi\epsilon_0rc}\cdot\frac{\sin{\theta}\cos{\theta}\hat{e}_x+\sin^2{\theta}\hat{e}_z}{1-\beta^2\cos^2{\theta}}\cdot \delta(t-(r\cos{\theta})/\beta c -r/c)
\end{align}
The ratio of peaks $E_{TR}/E_{direct}$ and time of peak field as a function of $r$ and $\theta$ register at the monitor $M$ is shown in Fig.~\ref{fig:DirectvsTR}. 
Since our frequencies of interest are far below the plasma frequency, $E_{TR}$ can be simply obtained by integrating over the frequency range of interest, inversely related to the bunch length. For $\sigma_t = 100$ ps bunch length, $d\omega = 1/2\pi\sigma_t$ and 
\begin{align}
    \frac{E_{direct,x}}{E_{TR,x}}= \frac{\pi\gamma c \sigma_t(1-\beta^2\cos^2{\theta})}{\beta r\sin^3{\theta}\cos{\theta}} \\
    \frac{E_{direct,z}}{E_{TR,z}}= \frac{\pi\gamma c \sigma_t(1-\beta^2\cos^2{\theta})}{\sqrt{2}\cdot 1.5^{1.5}\beta \cdot r\sin^4{\theta}}
\end{align}
and shortest time between the peaks of both components;
\begin{align}
    {t_{direct,x}}-{t_{TR,x}}&=  (r\cos{\theta})/\beta c -r/c\\
   t_{direct,z}-t_{TR,z}&= (r\cos{\theta})/\beta c -r/c + (R/(\sqrt{2}\gamma\beta c))
\end{align}

\subsection{Measured frequency response of Bikon Antenna}
\begin{figure}[H]
\centering 
\subfloat[]{\includegraphics[width=0.48\textwidth]{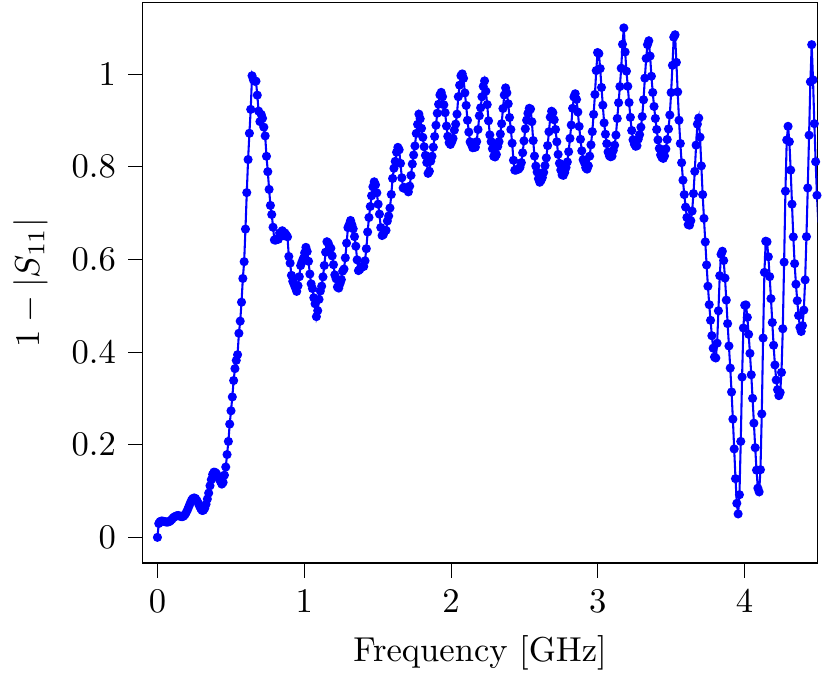}} 
\subfloat[]{\includegraphics[width=0.48\textwidth]{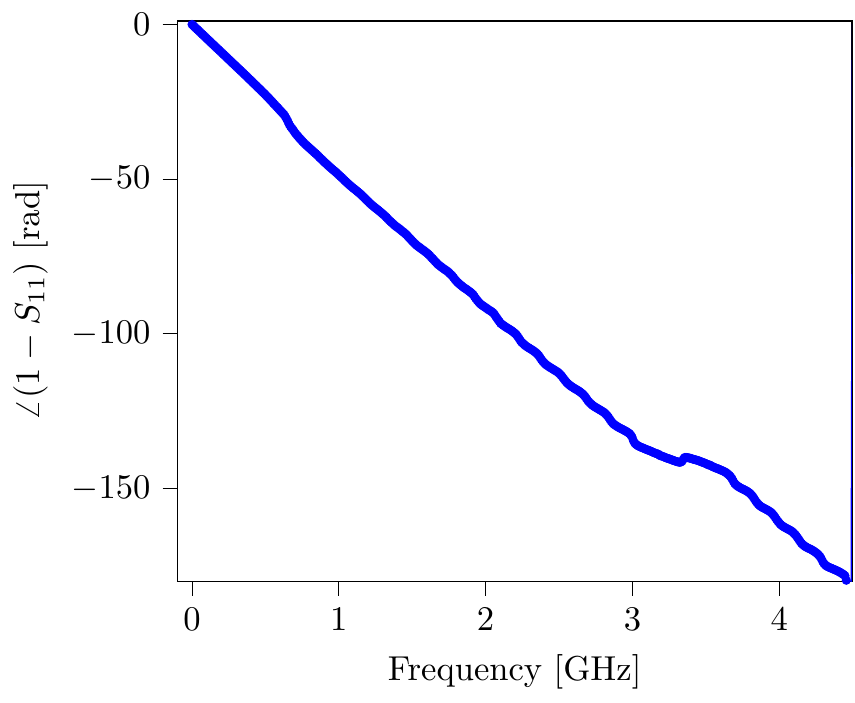}}
\caption{ (a) Magnitude response obtained using S11 measurement. (b) Phase response.}
\label{fig:bikon}
\end{figure}

Fig.~\ref{fig:bikon} shows the frequency response measured of the Biconical antenna used for these measurements. The measurements were not performed in an anechoic chamber. Further characteristics of the antenna can be found here~\cite{Schwarzbeck}.

\end{document}